\documentclass[fleqn,usenatbib,useAMS]{mnras}

\usepackage{amsmath}
\usepackage{graphicx}
\usepackage{hyperref}
\usepackage{siunitx}
\usepackage{booktabs}
\usepackage{natbib}
\usepackage[T1]{fontenc}
\usepackage{ae,aecompl}
\usepackage{newtxtext,newtxmath}

\DeclareSIUnit \parsec {pc}
\DeclareSIUnit \year {yr}
\DeclareSIUnit \mag {mag}
\DeclareSIUnit \msol {M_{\odot}}

\definecolor{orange}{rgb}{1.0,0.5,0.}

\def\MDM{\ifmmode{\>M_{\textnormal{\sc dm}}}\else{$$M_{\textnormal{\sc dm}}}\fi}

\def\XH{\ifmmode{\>X_{\textnormal{\sc h}}} \else{$X_{\textnormal{\sc h}}$}\fi}
\def\nH{\ifmmode{\>n_{\textnormal{\sc h}}} \else{$n_{\textnormal{\sc h}}$}\fi}

\def\maspyr{\ifmmode{\>\textnormal{mas~yr}^{-1}}\else{mas~yr$^{-1}$}\fi}

\def\mG{\ifmmode{\>\mu\mathrm{G}}\else{$\mu$G}\fi}
\def\erg{\ifmmode{\> {\rm erg}}\else{erg}\fi}
\def\keV{\ifmmode{\> {\rm keV}}\else{keV}\fi}

\def\deg{\ifmmode{\>^{\circ}}\else{$^{\circ}$}\fi}
\def\onedeg{\ifmmode{\>1^{\circ}}\else{$1^{\circ}$}\fi}

\def\xvir{\ifmmode{\>\!x_{vir}}\else{$x_{vir}$}\fi}
\def\Mvir{\ifmmode{\>\!M_{vir} }\else{$M_{vir} $}\fi}
\def\rvir{\ifmmode{\>\!r_{vir}}\else{$r_{vir}$}\fi}
\def\vvir{\ifmmode{\>\!v_{vir}}\else{$v_{vir}$}\fi}
\def\Vvir{\ifmmode{\>\!V_{vir} }\else{$V_{vir} $}\fi}

\def\tratio{\ifmmode{\>\tau}\else{$\tau$}\fi}

\def\rms{\ifmmode{\>r_{\textnormal{\sc ms}}}\else{$r_{\textnormal{\sc ms}}$}\fi}

\def\Mpc{\ifmmode{\>\!{\rm Mpc}} \else{Mpc}\fi}
\def\kpc{\ifmmode{\>\!{\rm kpc}} \else{kpc}\fi}
\def\pc{\ifmmode{\>\!{\rm pc}} \else{pc}\fi}

\def\Gyr{\ifmmode{\>\!{\rm Gyr}} \else{Gyr}\fi}
\def\Myr{\ifmmode{\>\!{\rm Myr}} \else{Myr}\fi}
\def\yr{\ifmmode{\>\!{\rm yr}} \else{yr}\fi}
\def\pyr{\ifmmode{\>\!{\rm yr}^{-1}}\else{yr $^{-1}$} \fi}
\def\s{\ifmmode{\>\!{\rm s}}\else{s}\fi}
\def\ps{\ifmmode{\>\!{\rm s}^{-1}}\else{s$^{-1}$}\fi}
\def\Hz{\ifmmode{\>\!{\rm Hz}}\else{Hz}\fi}

\def\kms{\ifmmode{\>\!{\rm km\,s}^{-1}}\else{km~s$^{-1}$}\fi}

\def\K{\ifmmode{\>\!{\rm K}}\else{K}\fi}

\def\sr{\ifmmode{\>\!{\rm sr}}\else{sr}\fi}
\def\psr{\ifmmode{\>\!{\rm sr}^{-1}}\else{sr$^{-1}$}\fi}
\def\arcs{\ifmmode{\>\!{\rm arcsec}}\else{arcsec}\fi}
\def\parcs{\ifmmode{\>\!{\rm arcsec}^{-1}}\else{arcsec${-1}$}\fi}
\def\parcss{\ifmmode{\>\!{\rm arcsec}^{-2}}\else{arcsec${-2}$}\fi}

\def\cm{\ifmmode{\>\!{\rm cm}}\else{cm}\fi}
\def\cc{\ifmmode{\>\!{\rm cm}^{3}}\else{cm$^{3}$}\fi}
\def\sqc{\ifmmode{\>\!{\rm cm}^{2}}\else{cm$^{2}$}\fi}
\def\pcc{\ifmmode{\>\!{\rm cm}^{-3}}\else{cm$^{-3}$}\fi}
\def\psc{\ifmmode{\>\!{\rm cm}^{-2}}\else{cm$^{-2}$}\fi}

\def\g{\ifmmode{\>\!{\rm g}}\else{g}\fi}
\def\Msun{\ifmmode{\>\!{\rm M}_{\odot}}\else{M$_{\odot}$}\fi}
\def\hMsun{\ifmmode{\> h^{-1}{\rm M}_{\odot}}\else{$h^{-1}$M$_{\odot}$}\fi}

\def\Zsun{\ifmmode{\>\!{\rm Z}_{\odot}}\else{Z$_{\odot}$}\fi}

\def\Lsun{\ifmmode{\>\!{\rm L}_{\odot}}\else{L$_{\odot}$}\fi}

\def\rayl{\ifmmode{\>\!{\rm R}}\else{R}\fi}
\def\mR{\ifmmode{\>\!{\rm mR}}\else{mR}\fi}

\renewcommand{\ion}[2]{\hbox{#1\,{\sc #2}}}

\def\lya{\ifmmode{\>\!{\rm Ly}\alpha}\else{Ly$\alpha$}\fi}

\def\Ha{\ifmmode{\>\!{\rm H}\alpha}\else{H$\alpha$}\fi}
\def\Hb{\ifmmode{\>\!{\rm H}\beta}\else{H$\beta$}\fi}

\def\HI{\ifmmode{\> \textnormal{\ion{H}{i}}} \else{\ion{H}{i}}\fi}
\def\HII{\ifmmode{\> \textnormal{\ion{H}{ii}}} \else{\ion{H}{ii}}\fi}
\def\CIV{\ifmmode{\> \textnormal{\ion{C}{iv}}} \else{\ion{C}{iv}}\fi}
\def\SiIV{\ifmmode{\> \textnormal{\ion{S}{iv}}} \else{\ion{Si}{iv}}\fi}

\def\NH{\ifmmode{\> {\rm N}_{\rm H}} \else{N$_{\rm H}$}\fi}
\def\Ng{\ifmmode{\> {\rm N}_{\rm gas}} \else{N$_{\rm gas}$}\fi}
\def\NHI{\ifmmode{\> {\rm N}_{\HI}} \else{N$_{\HI}$}\fi}
\def\MHI{\ifmmode{\> {\rm M}_{ \HI}} \else{M$_{\HI}$}\fi}

\def\mua{\ifmmode{\>\mu_{ \textnormal{\Ha}}}\else{$\mu_{ \textnormal{\Ha}}$}\fi}
\def\alphabha{\ifmmode{\>\alpha_{B}^{(\textnormal{\Ha})}}\else{$\alpha_{B}^{(\textnormal{\Ha})}$}\fi}

\newcommand{\gaia}{{\em Gaia}}

\newcommand{\lzvr}{$L_{z}-\bar{V}_{R}$\ }
\defcitealias{SgrImpactBHTG2021}{BT21}

\title[Effect of Single Impact on MW-like Disc]{
Galactic seismology: can a disc-crossing impulse explain the large-scale perturbations in the Milky Way's disc?}

\author[P. Yamsiri et al.]{
Pavadol Yamsiri,\(^{1}\)\thanks{Email: pyam4330@uni.sydney.edu.au}
Joss Bland-Hawthorn\(^{1}\) and
Thor Tepper-Garcia\(^{1}\)
\\
\(^{1}\)Sydney Institute for Astronomy, School of Physics, A28, The University of Sydney, Sydney, 
NSW 2006, Australia
}

\date{Last updated YYYY MM DD; in original form YYYY MM DD}

\pubyear{2026}

\begin{document}
\label{firstpage}
\pagerange{\pageref{firstpage}--\pageref{lastpage}}
\maketitle

\begin{abstract}
Prior to its infall, the Sagittarius (Sgr) dwarf galaxy was a major satellite with a mass of $M_{\rm sgr}\sim 10^{11}$ M$_\odot$. For the past $5-6$ Gyr, it has been heavily stripped by the Milky Way (MW), losing most of its mass while crossing the MW disc multiple times. Recent models of Milky Way disc perturbations $-$ including the spiral arms, the stellar bar, the \gaia\ phase spiral, and stellar and gaseous disc corrugations $-$ have identified these crossings as possible formation triggers, but have generally treated each perturbation in isolation. Here, we adopt a holistic perspective and ask whether a single disc-crossing impulse can simultaneously account for these features as observed today. We focus on simulations of single disc-crossing events by a Sgr-like perturber, and present a forensic analysis of the role of the powerful impulse in forming spiral arms, disc corrugations, the phase spiral and the `\lzvr wave', determined from a star's angular momentum and radial velocity, respectively. We find that a single disc crossing can reproduce reasonably well (e.g. structure, amplitude, phase) the observed local disc corrugation, and the Outer, Local and Sagittarius-Carina arm segments, implying that the last significant impulse due to a transit took place $700-1200~\Myr$ ago. Moreover, the \lzvr wave and phase spiral appear within the simulations over the same epoch and their general structure is reasonably well replicated, but not in detail. We conclude that Sgr's last significant crossing roughly a Gyr ago could be the primary cause of large-scale MW disc perturbations, but it cannot fully account for the \lzvr wave. Consequently, other triggers, possibly the Galactic bar or interactions with other satellites, must be considered in order to fully explain the current dynamical state of the MW's disc.
\end{abstract}


\section{Introduction}\label{sec:intro}

The advent of the European Space Agency (ESA) \gaia~astrometric mission and its major data
releases has led to the discovery of many phase space\footnote{In Galactic cylindrical coordinates ($R,\phi,z$), individual stars have velocities ($V_R$, $V_\phi$, $V_z$) and oscillation frequencies
($\Omega_R,\Omega_\phi,\Omega_z$) $=$ ($\kappa,\Omega,\nu$).} substructures in the Milky Way
stellar disc.
These features include ridges seen in $R-V_{\phi}$
\citep{Antoja2018,KawataRidges2018}, spirals in vertical phase space
$z-V_{z}$ known as \textit{phase spirals} \citep{Antoja2018}, planar waves
seen in $V_R$ \citep{FriskeLzVr2019} and many more. 
Such substructures are evidence that the Milky Way (MW) is in the process of settling from a
perturbed state in the recent past. In particular, the phase spiral, which appears as a
one-arm spiral pattern in the vertical phase-space distribution $z-V_z$ of stars in the local
Solar neighbourhood, is likely the result of phase mixing
\citep{LyndenBellPhaseMixing1967} after a strong impulsive disturbance of the disc.
Potential sources of these perturbations could be satellite galaxies like
Sagittarius (Sgr), bar buckling \citep{BarBuckling2019}, dark matter wakes induced by
satellites \citep{GrandDMWake2023}, or a series of stochastic kicks by objects like
molecular clouds or dark matter subhalos \citep{TremaineKicks2023}.

The MW disc has also been shown to be corrugated, displaying both small-scale, wave-like oscillations from the local disc
\citep{LocalVerticalWaveWidrow2012, LocalVerticalWavesGaiaDR2BennettBovy2019}
all the way to the outer disc, coexisting with the outer warp
\citep{OuterVerticalWavesSchoenrichDehnen2018}. In particular,
\citet{LocalVerticalWavesGaiaDR2BennettBovy2019} showed that the local vertical asymmetry
was found across various stellar colours and ages and that they were consistent,
indicating a single shared origin, such as from a single perturbation event \citep[see also][]{pog25}.
Corrugations are seen in high-resolution simulations of disc-satellite interactions \citep{PurcellSagittarius2011,Antoja2022TidalArms}. Notably, \citet{SgrImpactBHTG2021} first showed that an impulse-triggered corrugation (bending mode) winds up like the associated spiral density wave, but at half the angular rate.

Spiral arms are a common feature of almost all disc galaxies. The MW spiral arms have been observed across
different tracers, such as masers shown in studies
by \citet{ReidMasers2014} and \citet{SpiralArmMasersReid2019},
stars through surveys like \gaia~by \citet{PoggioGaiaSpirals2021}
and also in gas \citep[q.v.][]{SodingGasSpirals2025}.
For many decades, the origin of spiral arms and their exact nature
have been under constant debate with theories ranging from eternal and static density
waves \citep{LinShuStaticDensityWave1964}, now discredited, and transient structures popping in and out
of existence \citep{SellwoodCarlbergTransientSpirals1984}. Spiral arms can arise from
various mechanisms, including swing-amplified instabilities, bar-induced or tidally-induced excitations, or self-excited disc modes \citep{PurcellSagittarius2011,sell22}.

Thus the MW is host to many asymmetries that are plausibly explained by a tidal
interaction with an external perturber such as the Sgr dwarf. However it is still unclear how dominant are recent pericentric passage of Sgr to the current
structure and state of the MW. \citet{SgrImpactBHTG2021} have shown that a single
satellite disc crossing can incite both in-plane spiral density waves and vertical bending waves. These waves
do not only rotate but also wrap over time leading to increasingly wrapped spiral arms and corrugations respectively. The angular wrapping rates of such waves differ with
the bending wave wrapping slower than the spiral density wave, such that
spiral arms oscillate in height over azimuth as opposed to being confined to the
plane. This mechanism is proposed to be the most likely origin of the coupling between the vertical
phase spiral seen in mass density and its in-plane motion encoded by the mean radial
velocity \( \langle V_{R} \rangle \) and the mean azimuthal velocity
\( \langle V_{\phi} \rangle\).

This link and relationship between these tidally-induced spiral arms,
the disc corrugation and the phase spiral could serve as an explanation as to how these
features originated in the MW itself via a single shared origin.

This study aims to test this mechanism by further examining the N-body simulation 
used in \citet{SgrImpactBHTG2021}, along with more advanced models, to investigate the
plausibility of such an origin; that the last significant crossing of Sgr is responsible
for the current spiral structure, corrugated disc and phase spirals in the modern MW.
Included in our consideration is the
condition that such MW-like features must also be consistent with each other, meaning
that they are observed from the same point in time and space in the simulation.

The paper is set out in the following way: the experimental setup is given in Sec.~\ref{sec:setup},
the overall strategy in Sec.~\ref{sec:methods} followed by the results in
Sec.~\ref{sec:results}. Sec.~\ref{sec:discussion} will discuss how the addition
of gas and more advanced gas physics will affect our results and finally in
Sec.~\ref{sec:conclusion} we will summarise our findings.

\section{Experimental Setup}\label{sec:setup}
We evaluate the plausibility of such a model by looking at an isolated galactic disc
simulation, first presented in \citet{SgrImpactBHTG2021} (hereafter referred to as \citetalias{SgrImpactBHTG2021}).
In this model, the MW is approximated by a three component system,
which consists of: 1) a dark matter (DM) halo; 2) a pre-assembled stellar bulge; and 3) a pre-assembled stellar disc.
The three components are all responsive, and they are sampled with $2 \times 10^{7}$ particles (DM halo), and $4.5 \times 10^{6}$ particles (stellar bulge), and $5 \times 10^{7}$ particles (stellar disc). The masses of
the three components are approximately
$1.4 \times 10^{12}~\si{\msol}$, $1.5 \times 10^{10}~\si{\msol}$, and $3.4 \times 10^{10}~\si{\msol} $, respectively.

The simulations have been carried out with the {\sc Nexus} framework \citep{TTGJBHNexus2024}.
In brief, the initial conditions for each galaxy component were created using the Action-based Galaxy Modelling Architecture (AGAMA) stellar dynamics library \citep{VasilievAgama2019}, which have been complemented to include gas in addition to the standard treatment of collisionless components.

The initial conditions are evolved with the adaptive mesh refinement (AMR), N-body/hydrodynamical code {\sc Ramses} \citep{tey02a}, augmented with a proprietary module to account for galaxy formation physics \citep[q.v.][]{age21l}.

In order to keep the disc reasonably stable against the development of instabilities from the outset, we choose the stellar radial velocity dispersion $\sigma_{R}$ such that Toomre's stability parameter defined by
\begin{equation}
	Q_{*} \equiv \frac{\sigma_{R}\kappa}{3.36 G \Sigma_{*}}
\end{equation}
where $\kappa$ is the epicyclic frequency and $\Sigma_{*}$ is the stellar surface
density, is above $1.3$ across the disc. As a result of the the relatively low disc mass fraction, the model galaxy is stable against the bar-instability over at least $\SI{4}{\giga\year}$
\citep{FujiiBar2018, JBHTTGBarOnset2023}.

Originally, \citetalias{SgrImpactBHTG2021}'s model was designed to mimic the analytic toy
model presented by \citet{SgrImpactBS2018}, in which the Sagittarius dwarf galaxy (Sgr) is approximated by a point mass. Its orbital parameters are such that it intersects the galactic midplane of the disc at $R \approx 18~\kpc$  travelling at an infall speed of $330~\kms$.

In order to isolate the effects of a single
impact, the mass of the perturber is artificially reduced over time.
The mass evolves as follows
\begin{equation}
	M_P(t) = M_P(0) \times
	\left\{
	\begin{array}{ll}
		1 & t < t_{\circ} \\
		\exp\left(\frac{-(t - t_{\circ})}{\tau_{s}}\right) & t \geq t_{\circ}
	\end{array} 
	\right. 
\end{equation}
where $M_{P}(0) = 2 \times 10^{10}~\si{\msol}$ and
$\tau_{s} = \SI{30}{\mega\year}$, roughly a disc crossing time.
The perturber first crosses the disc at about $\sim \SI{100}{\mega\year}$ and so
$t_{\circ}$  is set to $\SI{150}{\mega\year}$.
This mass loss means that at the perturber's second disc crossing, its mass becomes
negligible.
Hence, this model serves as a useful proxy for exploring the effects of a
``Sagittarius dwarf''-sized impact onto the Milky Way disc.\footnote{We refer the reader to \citetalias{SgrImpactBHTG2021} for a more detailed account and description of the
simulation.}

\section{Methods}\label{sec:methods}
The main goal of this study is to explore whether a single, impulsive interaction between Sgr and the MW can account for the large-scale perturbation of the Milky Way disc we observe today.
We do this by comparing on a like-for-like basis the perturbations of the synthetic galaxy in \citetalias{SgrImpactBHTG2021}'s model with MW data, and quantifying their similarities and differences.
The details of our approach are presented and discussed in the following sections.

\subsection{Reference Frame}\label{sec:solar_circle}
All observations and measurements of the MW's structure are carried out in a heliocentric frame, while our simulations are effectively ran in a Galactocentric frame. A meaningful comparison between observations and simulations requires them to be in the same frame of reference; we choose for this purpose the heliocentric frame and thus map all of our simulation results to this frame, as explained next.

We adopt a Sun to Galactic centre distance of $R_0 = 8$~\kpc\ and a Sun to midplane height of $z_0 = 0$~\kpc\ to simplify the coordinate transformations.\footnote{This does make some comparisons with observations inconsistent as for example the spiral arms assume a different Galactocentric frame. This does not significantly change the results.}

In a synthetic MW disc featuring a central stellar bar, there are two possible and fully equivalent sun positions, each at a location $(R_0, z_0)$ and at an angle of roughly 28\deg\ with respect to either bar end \citep[see e.g.][their figure 3]{tep21v}. In contrast, in an perfectly axisymmetric disc - as is the case of our adopted model - there is no preference for any azimuth. Therefore, any location on a circle with radius equal to $R_0$ - the solar circle -- is a valid reference point to compare with observation, creating a circle of candidate reference frames.

\begin{figure}
\begin{center}
\includegraphics[width=\columnwidth]{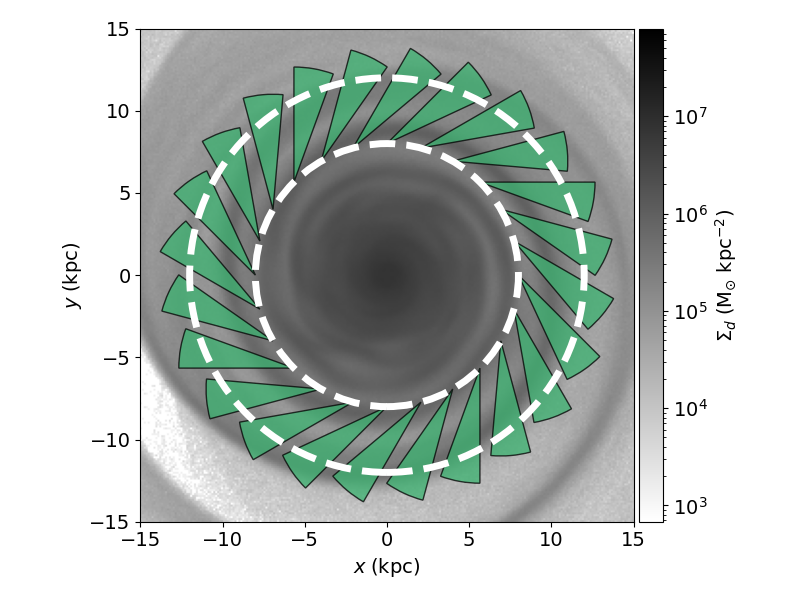}
\caption{Illustration of 24 wedge shaped selections shaded in green used to sample the radial
height profile or \textit{corrugation}, overlaid on top of a synthetic galactic disc that is rotating anti-clockwise.
In addition to the wedge shaped selection there are further cuts on 
Galactocentric radius denoted by the inner and outer dashed white circles. The selection is
an approximation of the projected selection used by \citet{CorrugationYH2024} in their analysis
of the local radial height profile.
Note that only 24 of the 72 wedges we used for our analysis are shown here to avoid excessive
visual clutter.
See Sec.~\ref{sec:method_corrugation} for a more detailed description of this selection.
\label{fig:solar_circle_demo}
}
\end{center}
\end{figure}

See Fig.~\ref{fig:solar_circle_demo} for a diagram of the reference points around the solar circle (inner white circle), and their respective selection regions (to be introduced in the following sections).

\subsection{General Strategy}\label{sec:general_strategy}

In order to quantify the correspondence between the simulated and the observed perturbations inflicted upon the MW's stellar disc we introduce hereafter a `similarity' metric individually for each corresponding feature: spiral arms, corrugation, phase spiral, and \lzvr wave

Given our choice of reference frame (Sec.~\ref{sec:solar_circle}), we apply each metric for 72 locations, i.e. azimuths, around the solar circle (cf. Fig.~\ref{fig:solar_circle_demo}, green shaded regions), and for a range of time steps over the full evolution of the system after impact.

For each feature, this results in a two-dimensional (2D) map in time (across snapshots) and in azimuth (along the solar circle) where the map value at each cell indicates the similarity between simulation and data, allowing for a visual assessment of the epoch and azimuth pairs at which the simulation provides a reasonable match to observations.

\subsection{Spiral Arms}\label{sec:method_spiral_arms}
\subsubsection{Observation}\label{sec:method_spiral_arms_observation}
To quantify the correspondence between simulation and observations with regards to the presence and structure of the stellar spiral arms, we use a spiral arm model adapted from
\citet[][see their Table~2]{SpiralArmMasersReid2019}.
The adopted models consist of arc segments {\em interpolated} across the locations of masers to limit the bias towards any particular spiral arm model. This means we do not adopt the
full spiral arm model as shown in Fig.~1 of \citet{SpiralArmMasersReid2019} as that involved
extrapolating fitted arm segments beyond their observed sources. Specifically, we consider the \textit{Scutum-Centauri}, \textit{Sagittarius-Carina}, \textit{Local}, \textit{Perseus} and \textit{Outer} arm segments, excluding the two innermost arms, the 3-kpc(N) and Norma 
arms appear too close to the Galactic bar, which
is missing in the simulations considered in this report.

The spiral arm segments are modelled as kinked
log spirals described by the following functional form
\begin{equation}
	\ln\left(\frac{R}{R_{\mathrm{kink}}}\right) =
		-\left(\beta - \beta_{\mathrm{kink}}\right) \tan{\psi_{\pm}}
\end{equation}
where $R$ is the Galactocentric radius and $\beta$ is the azimuth defined as
$0$ towards the Sun and increasing in the direction of Galactic rotation. The arm is
then parametrised by the following parameters: $R_{\mathrm{kink}}$ is the kink radius,
$\beta_{\mathrm{kink}}$ is the kink azimuth and $\psi_{\pm}$ are pitch angles
which can differ depending on whether $\beta$ is larger or smaller than $\beta_{\mathrm{kink}}$.

\begin{figure}
\begin{center}
\includegraphics[width=\columnwidth]{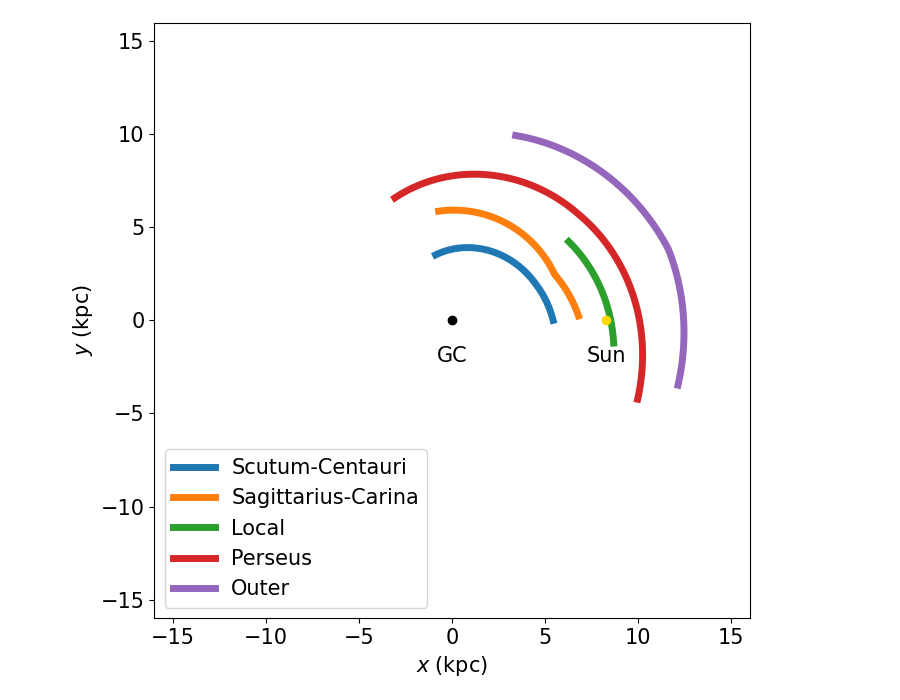}
\caption{Spiral arm segments adapted from \citet{SpiralArmMasersReid2019}, shown as they
would appear if the Sun were located at $(X,Y)=(8,0)~\kpc$ and the
disc rotated anti-clockwise.
Note that we do not use the four arm plus Local arm model proposed by
\citet{SpiralArmMasersReid2019} which extrapolated fitted arm segments instead plotting
the arm segments only across the azimuths where maser sources are observed using the
parameters seen in Table~2. The innermost arms, the $3~\kpc$(N) and Norma arms are excluded,
since they exist too close to the Galactic bar which is not included in the simulations
considered in this study.
\label{fig:spiral_arms}
}
\end{center}
\end{figure}

We refer the reader to Fig.~\ref{fig:spiral_arms} for a sketch of the adopted arm segments and their relative position to the Sun.

\subsubsection{Simulation}\label{sec:method_spiral_arms_simulation}
As explained in Sec.~\ref{sec:method_spiral_arms_observation}, our adopted
spiral arms are modelled in terms of a Galactocentric frame where
the Sun is assumed to be at $R = R_{\odot} = 8~\kpc, \phi = 0$ and $z = z_{\odot} = 0~\kpc$. 
Therefore, for each possible location along the solar circle we have a
continuous spectrum of possible positions of the spiral arms. We then need
to evaluate for each position how well the simulation reproduces the
MW spiral arms.

\subsubsection{Comparison}\label{sec:method_spiral_arms_comparison}
The similarity metric we will use for the spiral arms is based on a spatial `mask-based'
approach. In essence, we discretise the arms into pixel masks over a 2D Cartesian grid projection of the disc's stellar surface density in the $x-y$ plane.
Each mask identifies the region associated with a particular arm segment.
We compute the number of pixels within each mask that are classified as `overdense', and divide that by the total number of pixels contained within the mask. `Overdense'
cells are those with a higher stellar surface density relative to the initial
surface density when the disc was smooth, i.e. at the beginning ($t=0$) of the simulation.

The overdensity $\Delta$ is defined as follows
\begin{equation}
	\Delta(x, y, t) = \frac{\Sigma(x, y, t)}{\Sigma(x, y, 0)}
\end{equation}
where $\Sigma(x, y, z, t)$ is the stellar surface density at
point $x, y$ at time $t$. An overdense cell is then a cell where this
quantity is greater than $1$.

The spiral arm similarity metric is then defined as follows
\begin{equation}
	\mathcal{S}_{\text{arm}} =
	\frac
	{\# \, \text{of overdense cells within mask}}
	{\# \, \text{of total cells within mask}}
\end{equation}
This metric represents the fraction of overdense cells falling within a particular arm
segment region serving as a proxy for how well the simulated spiral arms are aligned
with the observed spiral arms. Its value ranges from $0$, which indicates poor
correspondence, to $1$, which implies perfect correspondence. We note that in principle
a value of $1$ does not actually mean perfect alignment as one can one can imagine
a disc with large overdense regions that fall within the masks but do not form spiral
arm segments resembling the given spiral arm models
(see Appendix~\ref{sec:spiral_arm_toy_model} for a more in-depth explanation).
However, for the simulations studied in this work, this effect is not that significant, as the overdensity map is a reasonable proxy for the locating spiral arms.

\subsection{Corrugation}\label{sec:method_corrugation}
Disc corrugations tend to span entire galactic discs \citep[see Fig. 21 in][]{bland19}. However, reliable measurements in \gaia\ trace only the local disc, in part due to dust
but also due to our position within the MW itself \citep{pog25}. Observationally,
we can really only sample part of the disc to a high degree of confidence. It is therefore not possible to compare the global structure of disc corrugations between
observations and simulations. Thus, we must settle for proxies of the global
disc corrugation in the form of local trends in the height of the midplane.

\subsubsection{Observation}\label{sec:method_corrugation_observation}
In a paper by \citet{CorrugationYH2024} they have shown the presence of a radial wave in
height in a wedge shaped region centred on the Sun.

Their selection can be summarised as follows:

\begin{enumerate}
\item{
	a cut in parallax $p \geq 0$
}
\item{
	a cut in $G$ band apparent magnitude $G_{\mathrm{mag}} \in [13, 18]~\si{\mag}$
}
\item{
	a cut in \gaia~colour $G_{\mathrm{BP}} - G_{\mathrm{RP}} \in [0.6. 2.4]$
}
\item{
	a cut in Galactic longitude $l \in [225, 245]~\si{\deg}$
}
\item{
	a cut\footnote{this cut was not mentioned in \citet{CorrugationYH2024}, however, it was used to obtain their results (confirmed via private communication).} given by
	\begin{equation}
		3.8857\: G_{\mathrm{BPRP}} - 0.4312
		\leq
		M_{G}
		\leq
		3.4251\: G_{\mathrm{BPRP}} + 1.517
	\end{equation}
    where
    \begin{equation}
    	M_{G} = G_{\mathrm{mag}} - 5\log_{10}\left(\frac{d}{\SI{10}{\parsec}}\right)
    \end{equation}
    is the absolute $G$ band magnitude and
    \begin{equation}
    	G_{\mathrm{BPRP}} = G_{\mathrm{BP}} - G_{\mathrm{RP}}
    \end{equation}
    is the difference between \( G \) band magnitudes.
}
\end{enumerate}
\
The selected stars are then binned radially from $8$ to $\SI{12}{\kilo\parsec}$
and the stars in each radial bin are then binned again in height. See the green shaded
regions in Fig.~\ref{fig:solar_circle_demo} for examples of the shape of the
selection.

The height distribution in each radial bin is passed through an odd low-pass filter to
remove any long wavelength odd oscillations and the mean height of the resulting
distribution is taken to be the mean height of the plane for that bin. The odd low pass
filter works as follows:
\begin{enumerate}
\item{
	Take the Fourier transform of a spatial distribution $\psi$
	\begin{equation}
		\psi(x) \to \psi(\tilde{\nu})
	\end{equation}
	where $x$ is the spatial coordinate and $\tilde{\nu}$ is the linear
	wavenumber.
}
\item{
	For wavenumbers greater than some set limit
	$\tilde{\nu} > \tilde{\nu}_{\mathrm{lim}}$, ignore the imaginary component or
	more specifically apply the following function $F(\tilde{\nu})$
	\begin{equation}
		F(\tilde{\nu}) \psi(\tilde{\nu}) = \left\{
		\begin{array}{ll}
		\psi(\tilde{\nu}) & \tilde{\nu} \leq \tilde{\nu}_{\mathrm{\lim}} \\
		\Re\left(\psi(\tilde{\nu})\right) & \tilde{\nu} > \tilde{\nu}_{\mathrm{\lim}}
		\end{array}
		\right.
	\end{equation}
	where $\Re(x)$ returns the real part of $x$.
}
\item{
	The resulting distribution in frequency space is converted back into the spatial
	domain to retrieve the filtered distribution.
}
\end{enumerate}
We adopt the same frequency cutoff $\tilde{\nu} = \SI{0.5}{\per\kilo\parsec}$ as
\citet{CorrugationYH2024}.

\subsubsection{Simulation}\label{sec:method_corrugation_simulation}
The selection function described in the previous section barring cuts on stellar
parameters like magnitude and colour can be replicated in our simulations allowing for a
direct comparison between the radial waves in the simulations and the wave observed by
\citet{CorrugationYH2024}.

In addition to the selection function we add uncertainty to the radial distances to each
star with respect to the reference point to mimic the observational distance errors and
make the comparison more fair.
The inclusion of distance errors is motivated by \citet{DistanceErrorHey2023} who have
demonstrated that uncertainty in distance can greatly influence observables.
For simplicity, distances are sampled from a distribution where the mean is its true
value and its half width at half maximum is equal to \( 15\% \) of its true value.

\subsubsection{Comparison}\label{sec:corrugation_metric}
We aim to show that galaxy models with a single impulse can reproduce the
corrugation. This requires a metric with which to quantify how much the simulation's
radial wave resembles the observations in order to conclude that the simulation can
``reproduce'' the corrugation.

There are a few properties we would like the metric to have:

\begin{enumerate}
\item{
	The value ranges from $0$ representing infinite dissimilarity and $1$
	representing that the two waves are identical.
}
\item{
	There is an intuitive interpretation of the values in between.
}
\end{enumerate}

Both criteria can be achieved if we have a method of calculating distance between the
simulated wave and the observed wave using the following equation
\begin{equation}
\text{similarity} = 1 - \frac{\text{distance}}{\text{distance} + \text{threshold}}
\label{eq:similarity_metric}
\end{equation}
where distance is any arbitrary quantity that ranges from $0$ meaning identical waves
and infinity meaning infinite dissimilarity and threshold refers to a threshold
distance with which we want to compare to. For example, the threshold could be the
distance between the observed wave and a wave resulting from a null model such as a
galaxy model without an impact.

The result of Eqn.~\ref{eq:similarity_metric} is a similarity metric where a value of
$0.5$ means the current test wave is as similar to the observed wave as the null
model. This then provides a natural threshold for determining if the simulations are
able to produce radial waves that better match the observed corrugation compared to
a null model. The null model in the case of the corrugation is that of a flat disc and
so we use the distance between the observed wave and a flat profile with height equal to
the mean height of the observed wave across all radial bins which is
$\sim \SI{8}{\parsec}$.

The distance score we will use from hereon is given by
\begin{equation}
	NSSR = \sum_{i=1}^{n} \frac{
		{\left(z_{i} - \hat{z}_{i}\right)}^{2}
	}{
		{\sigma_{z, i}}^{2} + {\sigma_{\hat{z}, i}}^{2}
	}
\end{equation}\label{eq:nssr}
which is simply the sum of square residuals each normalised by their combined variance.
See Appendix~\ref{appendix:residual_similarity} for a more in-depth discussion of how this
residual-based similarity metric is sensitive to the signals like the corrugation.

\section{Results}\label{sec:results}
\subsection{Spiral Arms}\label{sec:results_spiral_arms}
\begin{figure}
\begin{center}
\includegraphics[width=\columnwidth]{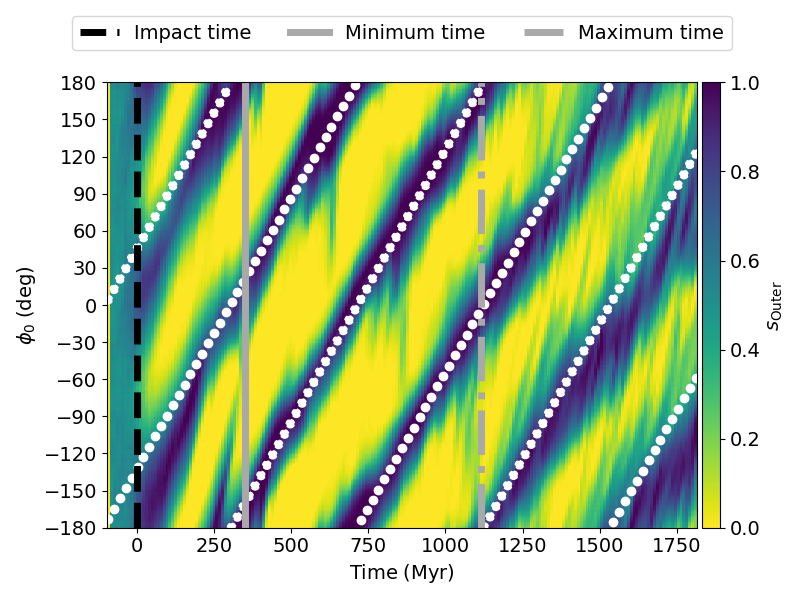}
\caption{The arm coverage similarity of the Outer arm. The time is set relative to the
time of impact and the azimuth is relative to the impact's azimuth, with positive
azimuths going in the anti-clockwise direction in the direction of Galactic rotation.
The impact time is outlined in black.
High arm coverage is coloured dark blue whereas low arm coverage is coloured in yellow.
The arm coverage is structured into diagonal bands that wrap in azimuth.
The minimum time and maximum time refer to the range in time where the bands appear
the most linear. A fit of the diagonal bands is also overlaid on top.
\label{fig:outer_arm_similarity}
}
\end{center}
\end{figure}

In order to simplify the analysis of the spiral arms, we will limit to considering
only the Outer arm at first. Using the mask-based similarity
(from here on called the arm coverage), we construct a map of the similarity across
time since impact and azimuth relative to the impact site's azimuth, which can be seen in
Fig.~\ref{fig:outer_arm_similarity}.

The arm coverage map displays remarkable distinct bands.
This indicates that the spiral arm segments in the simulation do resemble the
observed spiral arms and that they are rotating with the disc (see Appendix.~\ref{sec:spiral_arm_toy_model} for an explanation why).
Soon after the impact the bands are formed and become highly resolved aside from a few
kinks, but over time the bands' similarity starts to bleed out to other azimuths.

\begin{figure*}
\begin{center}
\includegraphics[width=2\columnwidth]{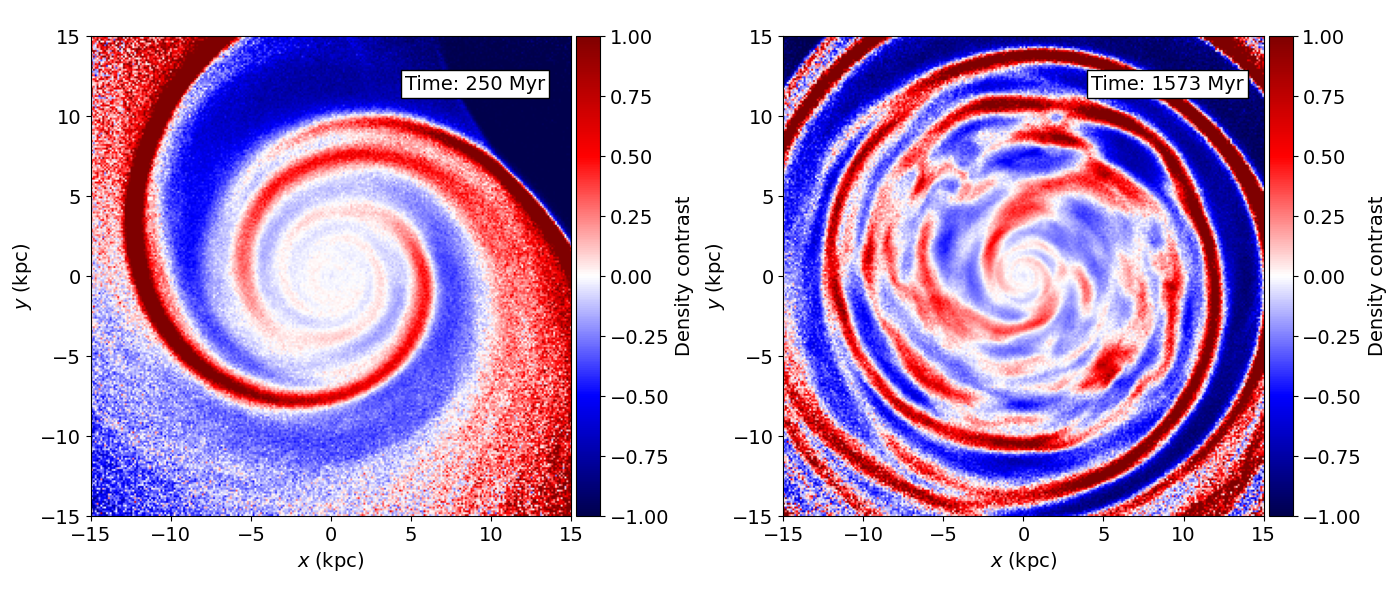}
\caption{Two snapshots of the density contrast at two different times.
Left: The density contrast at time $T = 250~\Myr$ where the impact has created
large spanning over and underdensity regions.
Right: The density contrast at a time late in the simulation where the disc has settled
leading to a weakening of the spiral arms which can be seen at late times in
Fig.~\ref{fig:outer_arm_similarity}.
\label{fig:hbd_7_hs4_wavefront}
}
\end{center}
\end{figure*}

This is a result of the impact inciting a large-scale density wave that rotates with
the disc. Over time the wave becomes increasingly wrapped from the centre outwards.
At late times, this wrapping is so severe it causes the spiral wave to decohere due to
diffusion between the wrappings of the arm, leading to the wispyness of the
high similarity bands in Fig.~\ref{fig:outer_arm_similarity} later on.
The evolution of the spiral density wave is illustrated in
Fig.~\ref{fig:hbd_7_hs4_wavefront}.

\begin{figure*}
\begin{center}
\includegraphics[width=2\columnwidth]{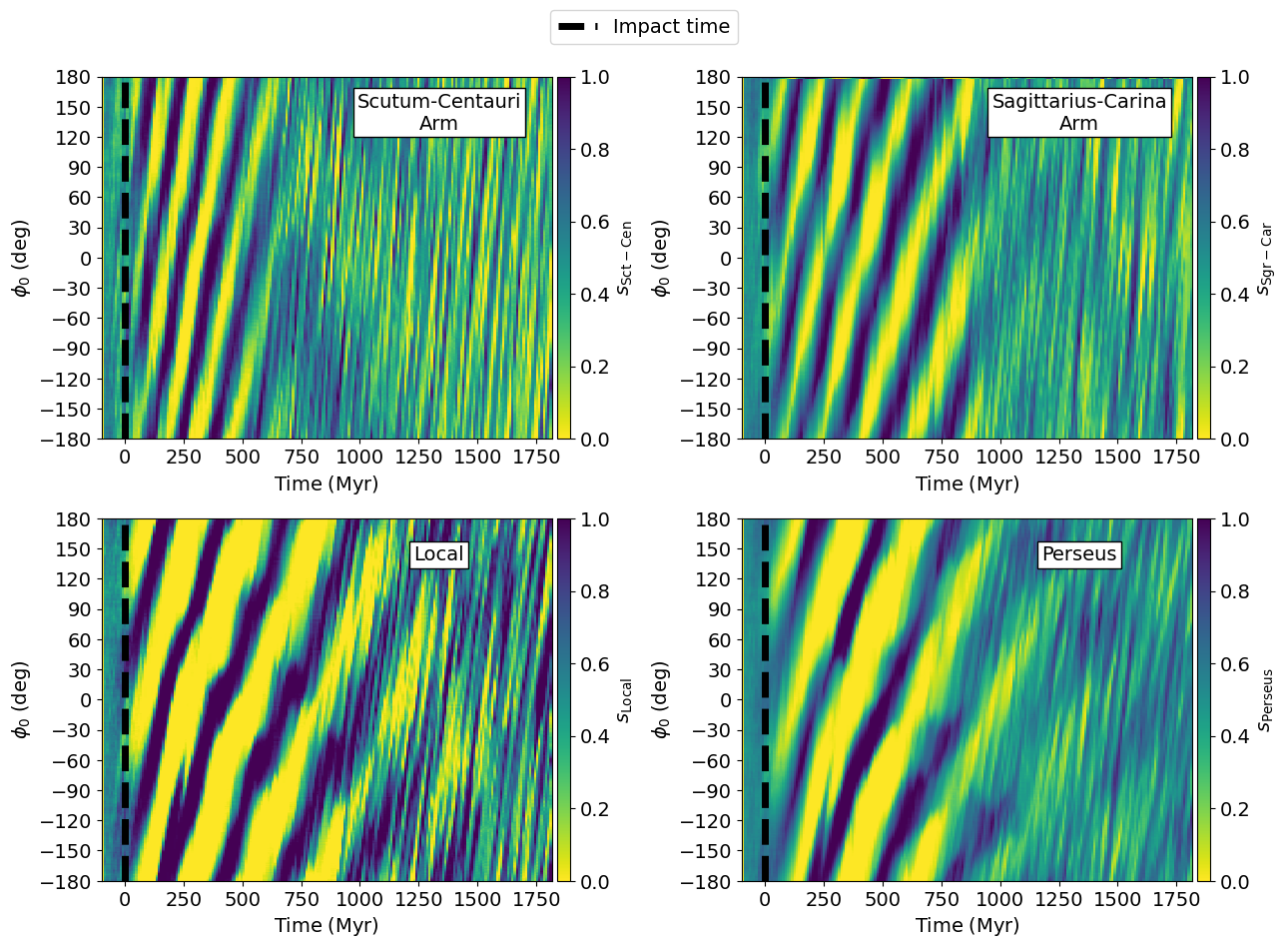}
\caption{The arm coverage similarity over azimuth and time for the Scutum-Centauri
arm (top left), Sagittarius-Carina arm (top right), Local arm (bottom left) and
the Perseus arm (bottom right). In each panel, the similarity traces out two separate wrapping
diagonal bands corresponding to an arm and its counter arm. These bands are highly resolved
some time after the impact but eventually dissipate as the disc settles, at varying rates
across arms.
\label{fig:hbd_7_hs4_other_arm_similarity}
}
\end{center}
\end{figure*}

The same method to create Fig.~\ref{fig:outer_arm_similarity} was used on the other four arm segments resulting in Fig.~\ref{fig:hbd_7_hs4_other_arm_similarity}. It is apparent that arm segments with smaller average radii (Scutum-Centauri and Sagittarius-Centauri) tend to have their similarity bands decohere faster than arm segments with larger average radii (Local, Perseus and Outer). This reflects that the spiral density wave diffuses from inside out.

\begin{figure}
\begin{center}
\includegraphics[width=\columnwidth]{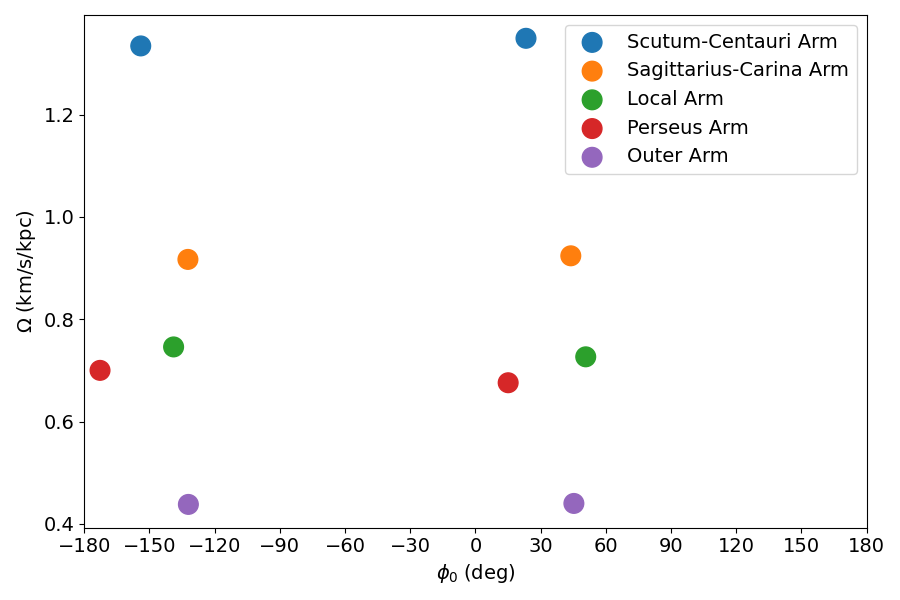}
\caption{The fitted parameters for the diagonal bands found in arm similarity.
Error bars are omitted as they are negligible at this scale.
\label{fig:ac_parameters}
}
\end{center}
\end{figure}

The diagonal bands can be modelled as rotating at a constant angular speed
\begin{equation}
	\phi(t) = \Omega t + \phi_{0}
\label{eq:linear_rotation}
\end{equation}
where $\Omega$ is the angular velocity in units of $\si{\deg\per\mega\year}$,
and $t$ is the time since impact in \Myr~and $\phi_0$ is an
initial phase offset in degrees. The bands in the similarity maps can be isolated by first ignoring all epoch-azimuth pairs with similarity
scores below $0.8$ and then collecting points that belong
to the same diagonal band into a set. Additionally, we limit the points to times where the bands appear the most linear visually, then for each set we 
apply linear regression assuming the model laid out in Eqn.~\ref{eq:linear_rotation} to obtain the band's angular speed and phase offset.
\
The parameters for the bands in arm coverage similarity
has been plotted in Fig.~\ref{fig:ac_parameters}.

It is clear from Fig.~\ref{fig:ac_parameters} that the
pattern speed appears to increase as the average radius of
the arm segments decreasing, which is unsurprising as the
angular speed increases towards the centre in a
differentially rotating disc.

The presence of high similarity diagonal bands for all
considered arm segments (see Fig.~\ref{fig:outer_arm_similarity} and Fig.~\ref{fig:hbd_7_hs4_other_arm_similarity}) demonstrates that the simulated
galaxy can achieve reasonable MW-like arm segments.

\subsection{Corrugation}\label{sec:results_corrugation}

\begin{figure*}
\begin{center}

\includegraphics[width=2\columnwidth]{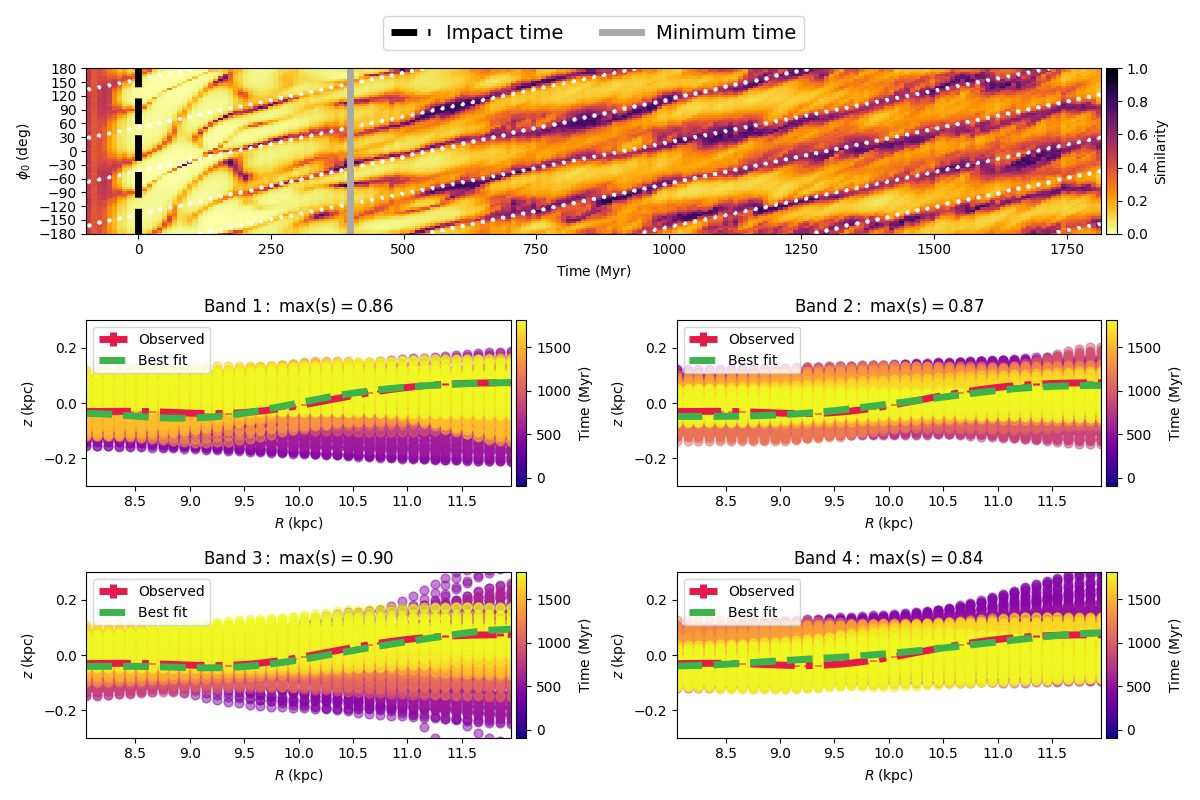}
\caption{
The top panel shows the similarity between the observed local radial height profile
and the simulated radial height profile or corrugation across the solar circle for all snapshots.
The time is set relative to the time of impact and the azimuth is relative to the impact's azimuth, with positive
azimuths going in the anti-clockwise direction in the direction of Galactic rotation.
The impact time is outlined in black.
High similarity is coloured dark purple whereas low arm coverage is coloured in yellow.
The similarity is structured into diagonal bands that wrap in azimuth like with arm coverage
similarity.
The minimum time refers to when the bands begin to appear and become linear. A fit of the diagonal bands is also overlaid on top.
The bottom panels show the radial waves over time along the four bands within $5\sigma$.
The maximum similarity along the bands are also displayed on top of each panel.
The time steps before the grey line at $T = 400~\Myr$ are ignored during
the fitting process and are also not displayed in the bottom panels. The best fitting
radial wave is also overlaid over the observed radial wave for each band.
\label{fig:corrugation_highlight}
}
\end{center}
\end{figure*}

We construct a map of the similarity between the simulation's corrugation and
the observed MW local corrugation using the method outlined in
Sec.~\ref{sec:corrugation_metric} which can be seen in
the topmost panel of Fig.~\ref{fig:corrugation_highlight}. We can see that the similarity starts off
relatively uniform across all azimuths and is $\sim 0.5$ as expected
since the disc is axisymmetric and stable before impact.
As the perturber approaches it starts to disrupt the disc causing the similarity to drop
drastically as the bending mode and breathing modes are excited, causing large scale
fluctuations which tend to fit poorly. There are however still azimuths where
the similarity does not change much which indicates that these exist at approximately the
nodes of the oscillations where the height profile does not drastically change.

After the perturber has crossed the disc, the similarity starts to increase as the
disc relaxes and flattens out. The excited bending and breathing modes are sheared by the
differential rotation of the disc leading to the formation of a corrugated disc.

\begin{figure}
\begin{center}
\includegraphics[width=\columnwidth]{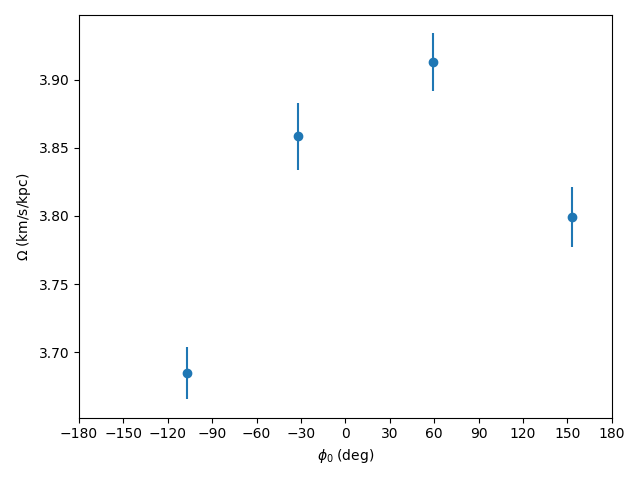}
\caption{
The fitted parameters for the four bands in the corrugation similarity map seen in
Fig.~\ref{fig:corrugation_highlight}. The error bars represent one standard deviation
$1\sigma$ of uncertainty.
\label{fig:cr_band_parameters}
}
\end{center}
\end{figure}

After about $400~\Myr$ after impact four distinct diagonal bands of high
similarity emerge. We can again model these using Eqn.~\ref{eq:linear_rotation},
and through the same process of manually masking the bands and applying linear regression
we were able to estimate values for the angular velocity $\Omega$ and offsets
$\phi_{0}$ for all four bands which can be seen in Fig.~\ref{fig:cr_band_parameters}.

The pattern speeds are similar to each other but not all of their error bars overlap.
The offsets are approximately equally spaced at ${90}^{\circ}$ apart forming a
four-fold symmetry.

The radial wave along these bands will not be exactly static but as the bands have
approximately constant similarity along it, this constrains the shape of the wave to at
least be within some bounds which can be seen in
Fig.~\ref{fig:corrugation_highlight}. All four bands achieve high similarity to the
observed radial wave at some point along it with values above $0.80$ and have relatively bounded radial profiles.

Overall, the simulation is able to produce convincing MW-like local disc corrugations.

\subsection{Estimating the impact time}\label{sec:results_time_estimate}
As the earlier sections show, the simulation is able to produce plausibly MW-like
observations of certain features at various azimuths and times; however, these features must also be consistent
with each other meaning they appear MW-like at the same solar azimuth and time. Applying this
constraint can lead to an estimate of the impact time of Sgr.

We apply the constraint of consistency by combining the similarity map of multiple features
into a single combined similarity map by applying the following procedure:
\begin{enumerate}
\item{
	Apply a threshold to each individual similarity map to mask out poor fits,
	specifically, $0.5$ for all residual based similarity maps like the corrugation
	similarity map, and $0.8$ for all arm coverage based similarity maps. These
	values were chosen as similarity scores above $0.5$ for residual based maps
	indicate better fits than the null model, whereas $0.8$ was chosen for arm
	coverage based similarity maps as it effectively masks out the bands in arm coverage
	similarity.
}
\item{
	Combine all the threshold similarity maps by computing their geometric mean.
}
\end{enumerate}

The resulting similarity map is the combined threshold similarity map where non-zero
values indicate where the fits for each included feature are individually above their respective
chosen thresholds.

First, we will consider the similarity maps for the corrugation and the spiral arms as they
fit the MW best. Both features contain high similarity bands
that are well approximated by Eqn.~\ref{eq:linear_rotation} and so it is expected that
in the combined similarity map, where these bands intersect will be regions of high
combined similarity. Therefore, by calculating their points of intersection, we can
estimate the time after impact and azimuth where the simulation displays the MW-like
features. The intersection between two bands characterised
by $(\Omega_{0}, \phi_{0})$ and $(\Omega_{1}, \phi_{1})$
is given by
\begin{equation}
	t_{\text{intersect},m} = \frac{\phi_{1} - \phi_{0} + 360m}{\Omega_{0} - \Omega_{1}}
\label{eq:intersection_time}
\end{equation}
where $t_{\text{intersect},m}$ is the intersection time
and $m$ is some integer. The azimuth at that time of intersection is then simply
\begin{equation}
	\phi_{\text{intersect},m} = \Omega_{0}
	t_{\text{intersect},m}
	+ \phi_{0}.
\label{eq:intersection_azimuth}
\end{equation}
See Appendix~\ref{appendix:intersection_point_derivation} for the full derivation of Eqn.~\ref{eq:intersection_time}
and Eqn.~\ref{eq:intersection_azimuth}, along with the
equations to compute the variance of the intersection point.

\begin{figure}
\begin{center}
\includegraphics[width=\columnwidth]{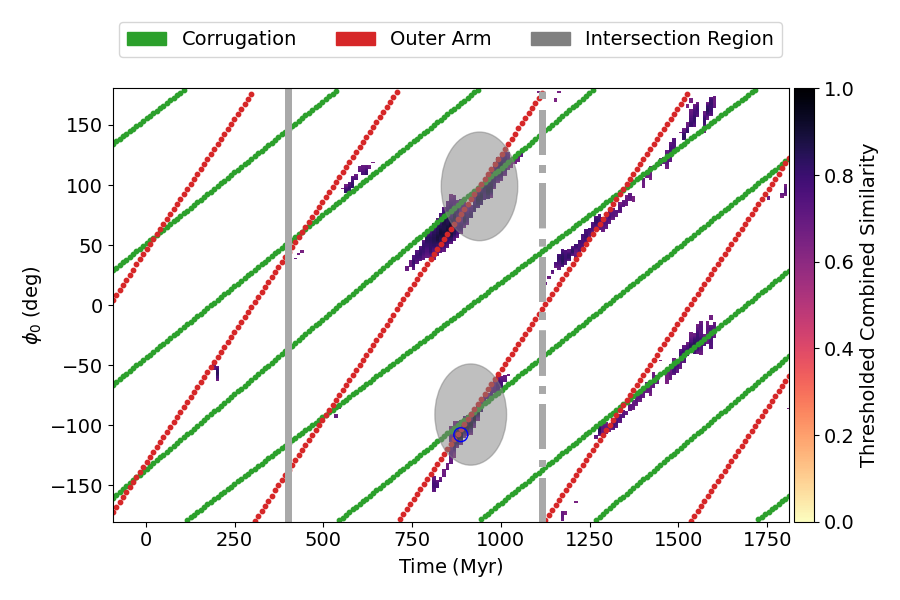}
\caption{A plot of the threshold combined similarity of the Outer arm similarity and
corrugation similarity overlaid with linear fits to each of their bands. The
$5\sigma$ confidence regions corresponding to the estimated point of intersections are highlighted as grey regions. The minimum and
maximum time bounds are denoted by the grey vertical lines and were decided by manual
visual inspection of where the bands are most linear.
The thresholded combined similarity is equal to the geometric mean of the Outer arm
similarity and the corrugation similarity wherein similarities below $0.8$
and $0.5$ for arm similarity and corrugation similarity are respectively culled (coloured in
white).
Intersection regions with too few cells with high combined similarity are discarded.
\label{fig:corrugation_vs_outer}
}
\end{center}
\end{figure}

Combining the Outer arm's similarity map and the corrugation similarity map results in the map
shown in Fig.~\ref{fig:corrugation_vs_outer}. From the map, we see that there are regions of high combined similarity near to the intersection between the high similarity bands, as suspected. The points of intersection for high combined similarity are
$T = 940 \pm 20 \, \Myr, \phi = 100 \pm 8 \, \si{\degree}$ and
$T = 915 \pm 19 \, \Myr, \phi = -91 \pm 8 \, \si{\degree}$.

\begin{figure*}
\begin{center}
\includegraphics[width=2\columnwidth]{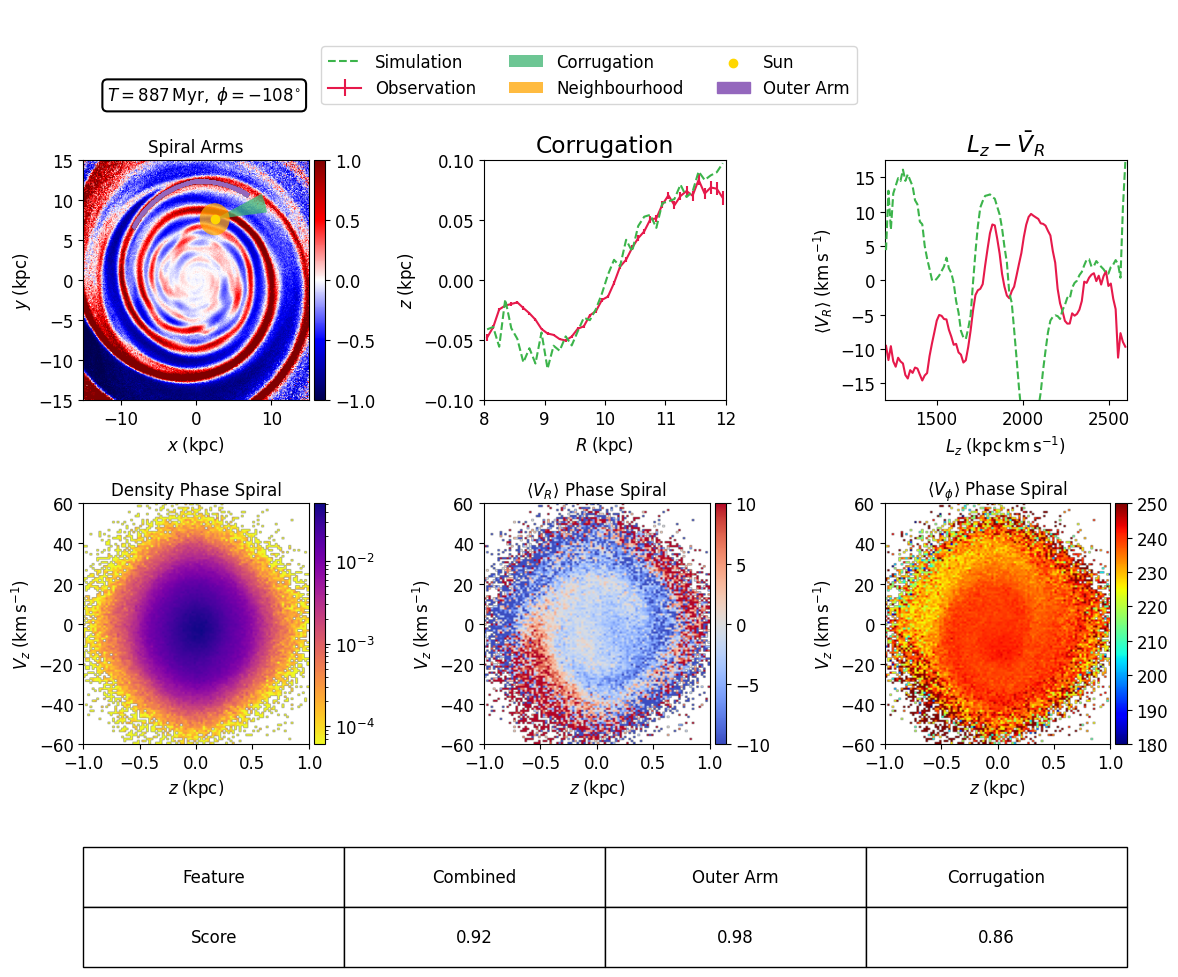}
\caption{
Top left: Density contrast map of the simulated disc galaxy. The model Outer arm is overlaid
in purple, the Sun’s position is marked in yellow, the neighbourhood used for the phase spiral is shaded in yellow, and the region used to extract the
corrugation profile is shown in green.
Top middle: Radial height profile of the disc corrugation, comparing the observed MW
profile (red) with the simulated profile (green).
Top right: The mean radial velocity binned in angular momentum or the \lzvr  wave, comparing
the observed MW profile (red) with the simulated profile (green).
Bottom left: The simulated vertical phase space distribution seen in surface number density.
Bottom middle: The simulated vertical phase space distribution seen in mean radial velocity (in units of $\kms$).
Bottom right: The simulated  vertical phase space distribution seen in mean azimuthal velocity (in units of $\kms$).
All non-observational data correspond to the simulation snapshot at $T = 887~\Myr$ and
solar azimuth $\phi = \SI{-108}{\degree}$, the configuration that maximises the
combined similarity score of the Outer arm and corrugation. A table of the similarity scores
for the Outer arm and corrugation is located below the panels.
\label{fig:corrugation_vs_outer_highlight}
}
\end{center}
\end{figure*}

As can be seen from Fig.~\ref{fig:corrugation_vs_outer},
the corrugation and the Outer spiral arm do seem to appear to match well at the
same time at around $T = 750-1200 \, \Myr$ and at two opposing sides
of the disc. Assuming the corrugation and Outer
arm segment in the MW are caused by an impact by Sgr, then we estimate that such an impact occurred
$\sim 1$ Gyr ago. This is confirmed in Fig.~\ref{fig:corrugation_vs_outer_highlight} that shows the simulated galaxy at the specific azimuth and time that maximises the combined similarity.

Including the other spiral arm segments is also possible, but as each arm segment has high
similarity bands with differing pattern speeds (as seen in
Fig.~\ref{fig:ac_parameters}), it is unlikely for all bands to intersect
at a single point like in the case of two features. Although, as the bands have a non-zero width it is still possible for them to appear MW-like at the same time.

\begin{figure*}
\begin{center}
\includegraphics[width=2\columnwidth]{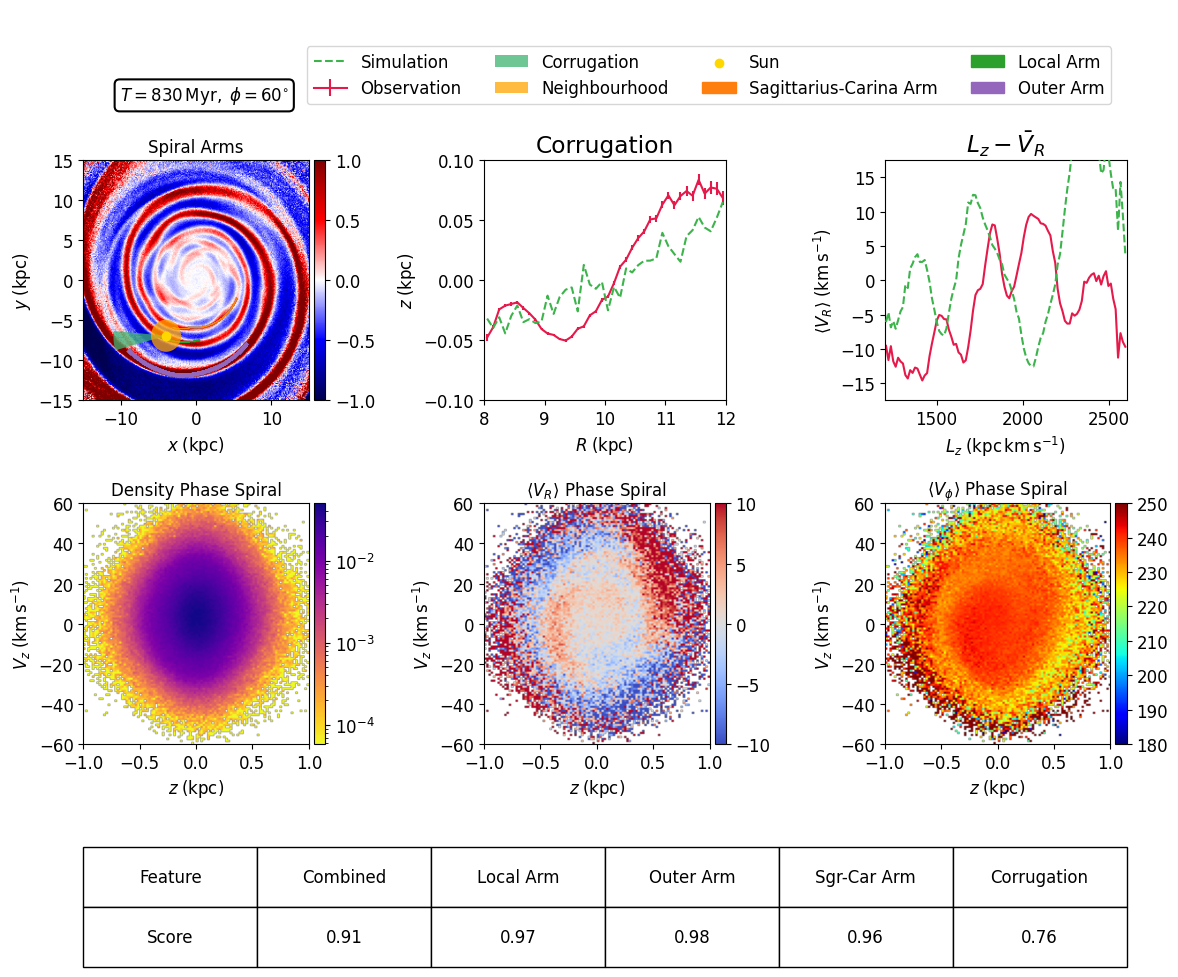}
\caption{
Top left: Density contrast map of the simulated disc galaxy. The model Outer arm is overlaid
in purple, the model Local arm is coloured in green, the model Sagittarius-Carina arm is in orange, the Sun’s position is marked in yellow, the neighbourhood used for the phase spiral is shaded in yellow and the region used to extract the
corrugation profile is shown in green.
Top middle: Radial height profile of the disc corrugation, comparing the observed MW
profile (red) with the simulated profile (green).
Top right: The mean radial velocity binned in angular momentum or the \lzvr  wave, comparing
the observed MW profile (red) with the simulated profile (green).
Bottom left: The simulated vertical phase space distribution seen in surface number density.
Bottom middle: The simulated vertical phase space distribution seen in mean radial velocity (in units of $\kms$).
Bottom right: The simulated  vertical phase space distribution seen in mean azimuthal velocity (in units of $\kms$).
All non-observational data correspond to the simulation snapshot at $T = 830~\Myr$ and
solar azimuth $\phi = \SI{60}{\degree}$, the configuration that maximises the
combined similarity score of the Outer arm, Local arm, Sagittarius-Carina arm and corrugation.
A table of the similarity scores for the corrugation and the chosen spiral arms is located below the panels.
\label{fig:hbd_7_hs4_compatible_arms}
}
\end{center}
\end{figure*}

We do not find any specific azimuth and time such that all arm segments and the corrugation match
well at the same time; however, after manual testing, the largest set of compatible segments
are the Outer, Local, Sagittarius-Carina arms and the local disc corrugation. See
Fig.~\ref{fig:hbd_7_hs4_compatible_arms} for a plot of the snapshot and azimuth which maximises
their combined similarity. Note that the inclusion of these other arms leads to a worse match between the simulated corrugation and the observed corrugation, however, the general trend of increasing height towards outer radii is preserved.

We conclude that a Sgr-like impact, while capable of producing some MW-like spiral arms and corrugation, is unlikely to be the sole mechanism that triggered all arm segments.
Specifically, the simulation does not seem to be able to reproduce the other arm segments like
the Perseus arm or any of the inner spiral arm segments like the Norma arm at the same time
as the other arm segments. A possible explanation is the non-linear behaviour seen at late times due to the tight-winding, inter-arm interaction.

\section{Discussion}\label{sec:discussion}

\subsection{Phase Spiral}\label{sec:results_phase_spiral}
The phase spiral is a clear sign of a past perturbation, and so it is not unreasonable, with hindsight, to expect spiral-like patterns to occur in the vertical phase space in our simulated galaxy, as many studies have now verified \citep[q.v.][]{hunt25}. However, none of these important works considered the impact of gas.
Recent work by \citet{ThorPhaseSpiral2025} on this particular simulated galaxy model has shown
that phase spirals are detectable from $400~\Myr$ to $1200~\Myr$ after impact. This result
is consistent with the timescales of crossings between the Outer arm
similarity and corrugation similarity that we found in Sec.~\ref{sec:results_time_estimate}.

Phase spirals are also present at the crossings between similarity bands
as well which can be seen in the bottom panels of Fig.~\ref{fig:corrugation_vs_outer_highlight} and Fig.~\ref{fig:hbd_7_hs4_compatible_arms}, where Fig.~\ref{fig:corrugation_vs_outer_highlight} is the
epoch and azimuth that maximises the combined similarity of the Outer
arm and corrugation whereas Fig.~\ref{fig:hbd_7_hs4_compatible_arms}
maximises the combined similarity of the Local, Outer, Sagittarius-Carina arms and the corrugation. The stars comprising the phase spirals in these figures are those enclosed
by a sphere of radius $1~\kpc$ centred on each location along the solar circle.

We will defer the quantitative analysis of the degree of similarity
between the simulated phase spiral and observed \gaia\ phase spiral
to future work as its complex phase space structure is harder to categorise.

\subsection{\texorpdfstring{\lzvr wave}{Lz-VR wave}}\label{sec:discussion_lzvr}
So far we have only considered the spiral arms, the corrugation and the phase spiral, extending the work of an earlier study \citep{SgrImpactBHTG2021}.
But there are also other MW features
indicative of a past perturbation. 
The \lzvr wave is among many features seen in the \gaia~data, first discovered by
\citet{FriskeLzVr2019}. It appears as an in-plane wave pattern in the mean radial velocity
$\langle V_{R} \rangle$ when binned in angular momentum $L_{z}$.
Like the other features discussed, we suspect this pattern is a signal, at least in part, of a past perturbation.

We sample a sphere centred on the Sun of radius $\SI{2}{\kilo\parsec}$ and then
bin in angular momentum from $\SI{1200}{\kilo\parsec\kilo\metre\per\second}$
to $\SI{2600}{\kilo\parsec\kilo\metre\per\second}$; we then calculate the mean
radial velocity of the stars in each bin. The observed sample is from \gaia~DR2 using the
same cuts used by \citet{FriskeLzVr2019}.

\begin{figure}
\begin{center}
\includegraphics[width=\columnwidth]{
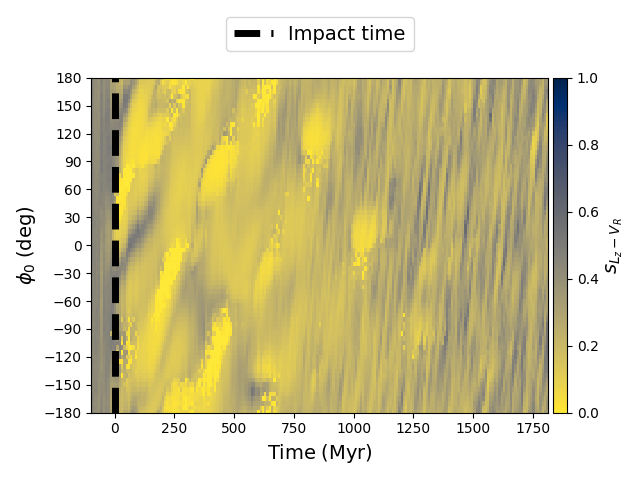
}
\caption{The similarity of the \lzvr wave as a function of azimuth and
time where time is set relative to the time of impact and the azimuth is relative to
the impact's azimuth, with positive azimuths going in the anti-clockwise direction.
There are distinct regions of low similarity (coloured in yellow).
\label{fig:hbd_7_hs4_lzvr}
}
\end{center}
\end{figure}

We attempted to find an epoch-azimuth pair that agreed well with the observed \lzvr wave using residual-based similarity (previously used to quantify corrugation similarity). Fig.~\ref{fig:hbd_7_hs4_lzvr} shows our similarity map for the \lzvr wave and it clearly lacks the bands of high
similarity found in the spiral arms and corrugation maps. Thus, for this diagnostic, we fail to find a convincing match.

This similarity map is very different when compared to the corrugation
and spiral arm maps. Rather than correlated band features, we see regions of low and medium similarity
instead. The simulation does not produce any reasonably MW-like waves in \lzvr~space except stochastically at late
times, $T > 1250~\Myr$, with the similarity averaging around $0.2$. The \lzvr projection at the crossings shown in Fig.~\ref{fig:corrugation_vs_outer_highlight} and Fig.~\ref{fig:hbd_7_hs4_compatible_arms} in the top right panels is also not convincingly MW-like, with structural differences in the simulated (green) pattern compared to the observed pattern (red).

This can be attributed to the more complex structure of the
observed \lzvr~wave that contains at least two dominant wave components compared
to the one wave component of the corrugation with shorter wavelengths relative to the
observed window. Additionally, previous work by \citet{BernetSindy2025} predicts that the amplitude ratio between the fast and slow waves in the \lzvr~signature
does not support a satellite impact origin, but rather points towards an internal mechanism.
\citet{CaoLzVr2024} also show evidence against an external cause as well. Hence, it seems unlikely the \lzvr wave originates from a single Sgr-like impact as confirmed by our results.

\subsection{Inclusion of Gas}\label{sec:discussion_gas_inclusion}
The simulated galaxy model so far lacks important physics, e.g., the  presence of a gaseous component.
In the following section, we explore how adding inert gas and star-forming gas can affect the formation of structures by applying the same methodology as described earlier
to these models \citep[q.v.][]{TTGJBHNexus2024}.

The model studied so far has been a pure N-body realisation. As \citet{ThorPhaseSpiral2025}
shows, the inclusion of gas can greatly influence the evolution and formation of features
such as the phase spiral, hence it is also important to consider its effects here. The
following sections will explore two more models which include gas and more realistic
physics and see how these extra inclusions affect the similarity.

\subsubsection{Inert Gas Model}\label{sec:discussion_inert_gas}
We will be utilising an inert isothermal gas galaxy model first presented in \citet{TGBHSgrGas2022}.
This model is exactly the same as the model used above; however, there is
an added component comprising an inert isothermal gas, distributed as an exponential disc
with scale length $R_{d} = 6~\kpc$ with mass $M_{\text{gas}} \approx 4\times 10^9~\si{\msol}$,
with a $10\%$ gas fraction relative to the total disc mass.
The gas is evolved via a strict isothermal equation of state with temperature $T = 10^3~\si{K}$ and is thus inert. See Sec.~2 of \citet{ThorPhaseSpiral2025} for a more in-depth
explanation of the model's parameters and setup labelled \textit{fg10\_nsf}.

\begin{figure}
\begin{center}
\includegraphics[width=\columnwidth]{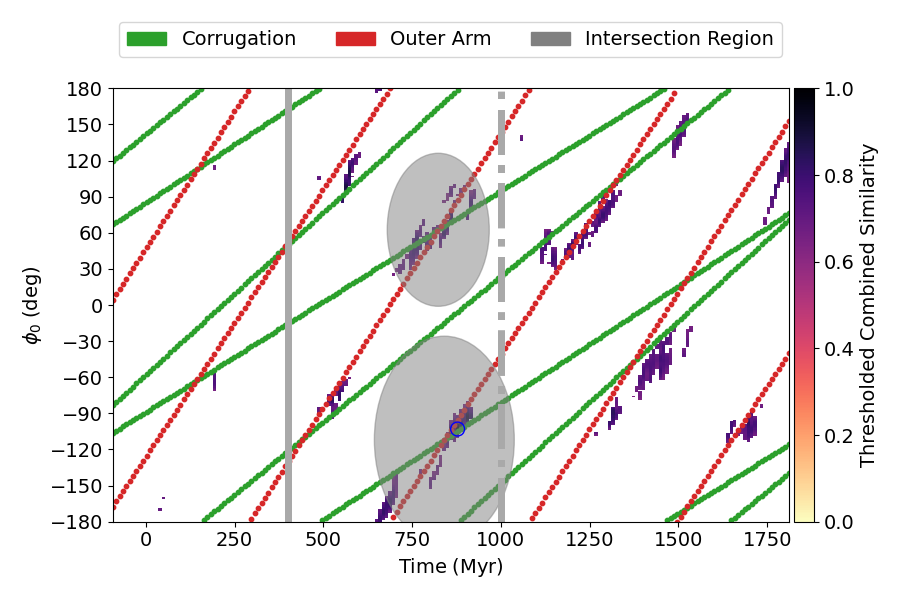}
\caption{Same plot as in Fig.~\ref{fig:corrugation_vs_outer} but of the inert isothermal
gas galaxy model. The blue circle corresponds to the best matching time and
azimuth with respect to the combined similarity of the Outer arm and corrugation.
\label{fig:hbd_10_gd2_intersection}
}
\end{center}
\end{figure}

We apply the same methods as laid out in Sec.~\ref{sec:methods} to these models to plot the
combined similarity map of the Outer arm similarity and corrugation similarity to produce
Fig.~\ref{fig:hbd_10_gd2_intersection}.
As compared to the gasless model, the regions
are much larger due to the increased uncertainty in the fitted parameters. This is a
result of the model increasing the noise in the similarity maps making the bands hazier.
Overall, the inclusion of an inert gas disc appears to reduce the times and azimuths where
multiple features match well at the same time.

\begin{figure*}
\begin{center}
\includegraphics[width=2\columnwidth]{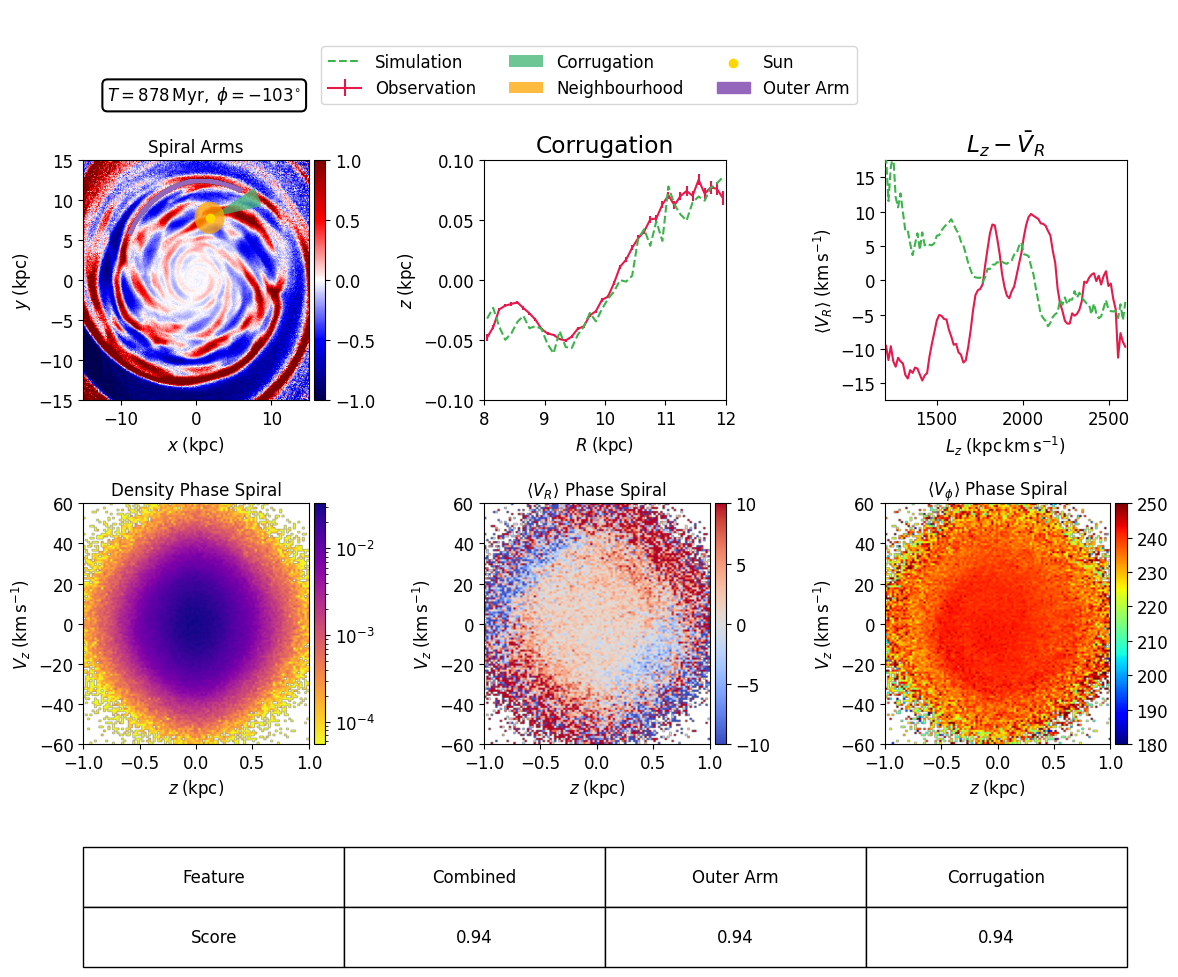}
\caption{
Top left: Density contrast map of the simulated disc galaxy with an \textbf{inert gas} disc. The model Outer arm is overlaid
in purple, the Sun’s position is marked in yellow, the neighbourhood used for the phase spiral is shaded in yellow, and the region used to extract the
corrugation profile is shown in green.
Top middle: Radial height profile of the disc corrugation, comparing the observed MW
profile (red) with the simulated profile (green).
Top right: The mean radial velocity binned in angular momentum or the \lzvr  wave, comparing
the observed MW profile (red) with the simulated profile (green).
Bottom left: The simulated vertical phase space distribution seen in surface number density.
Bottom middle: The simulated vertical phase space distribution seen in mean radial velocity (in units of $\kms$).
Bottom right: The simulated  vertical phase space distribution seen in mean azimuthal velocity (in units of $\kms$).
All non-observational data correspond to the simulation snapshot at $T = 878~\Myr$ and
solar azimuth $\phi = -\SI{103}{\degree}$, the configuration that maximises the
combined similarity score of the Outer arm and corrugation.
A table of the similarity scores for the corrugation and the Outer arm is located below the panels.
\label{fig:hbd_10_gd2_highlight}
}
\end{center}
\end{figure*}

A plot of the best fit epoch and azimuth for the inert gas model is also presented in Fig.~\ref{fig:hbd_10_gd2_highlight}. It is apparent from the top left panel that the spiral
arms are much more flocculent than in the pure N-body case presented in Fig.~\ref{fig:corrugation_vs_outer_highlight} and Fig.~\ref{fig:hbd_7_hs4_compatible_arms}.
The phase spirals in the bottom panels are also much weaker in comparison. Finally, the projection \lzvr does not match the observations at all.

\subsubsection{Star Forming Gas Model}\label{sec:discussion_active_gas}
In the previous section we included a gas disc; however, that gas was isothermal and inert, therefore, we have not included any form of stellar feedback or turbulence. 
In this section, we now introduce multi-phase gas that can undergo star formation leading
to stellar feedback. The model we will studying was first presented in \citet{ThorPhaseSpiral2025}, specifically, the \textit{fg20\_sf} model which in addition to the stellar components of
the pure N-body model also includes a gas disc with mass equal to 20\% of the total disc mass,
so $M_{\text{gas}} \approx 9 \times 10^{9}~\si{\msol}$ and with an initial temperature of
$T = 2\times 10^3~\si{K}$. We refer the reader to Sec.~2 of \citet{ThorPhaseSpiral2025} for a more
in-depth explanation of the simulation parameters.

As the gas can now form stars, this creates a distinction
between stars that are pre-existing and stars that are newly formed. In order to keep
the comparisons fair, we elect to only consider the pre-existing stars in this study.

\begin{figure}
\begin{center}
\includegraphics[width=\columnwidth]{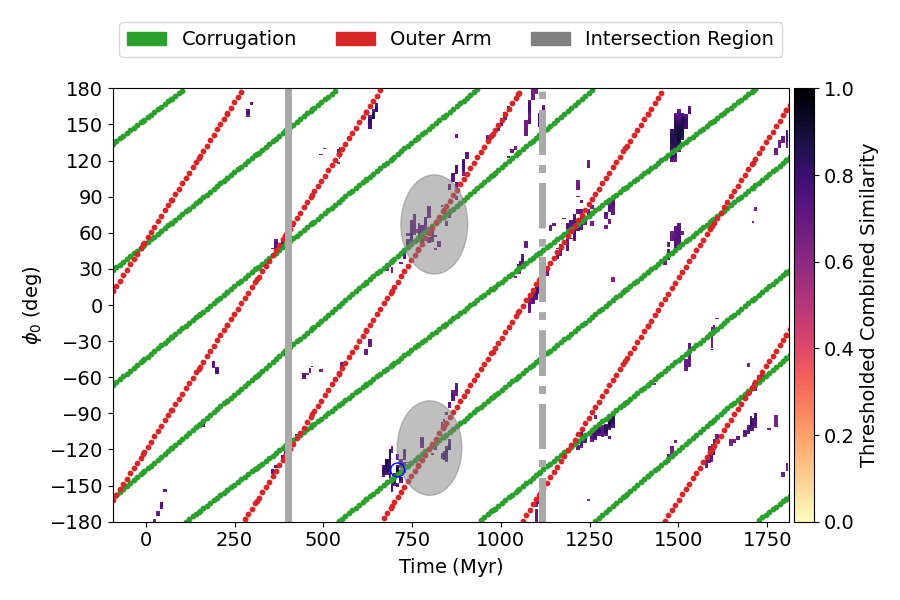}
\caption{Same plot as in Fig.~\ref{fig:corrugation_vs_outer} but of the multi-phase
star forming gas galaxy model. The blue circle corresponds to the best matching time and
azimuth with respect to the combined similarity of the Outer arm and corrugation.
\label{fig:hbd_10_gd11_intersection}
}
\end{center}
\end{figure}

We apply the same methods as in Sec.~\ref{sec:discussion_inert_gas} to obtain an intersection plot showcasing
the combined similarity of the Outer arm and corrugation which can be seen in
Fig.~\ref{fig:hbd_10_gd11_intersection}. Although the regions are  more
constrained, they contain less cells with non-zero threshold combined similarity 
compared to previous models. The introduction of feedback increases the speed of
decoherence in the spiral arms, and the added energy causes the radial wave oscillations
to be messier, resulting in even more flocculent similarity maps.

\begin{figure*}
\begin{center}
\includegraphics[width=2\columnwidth]{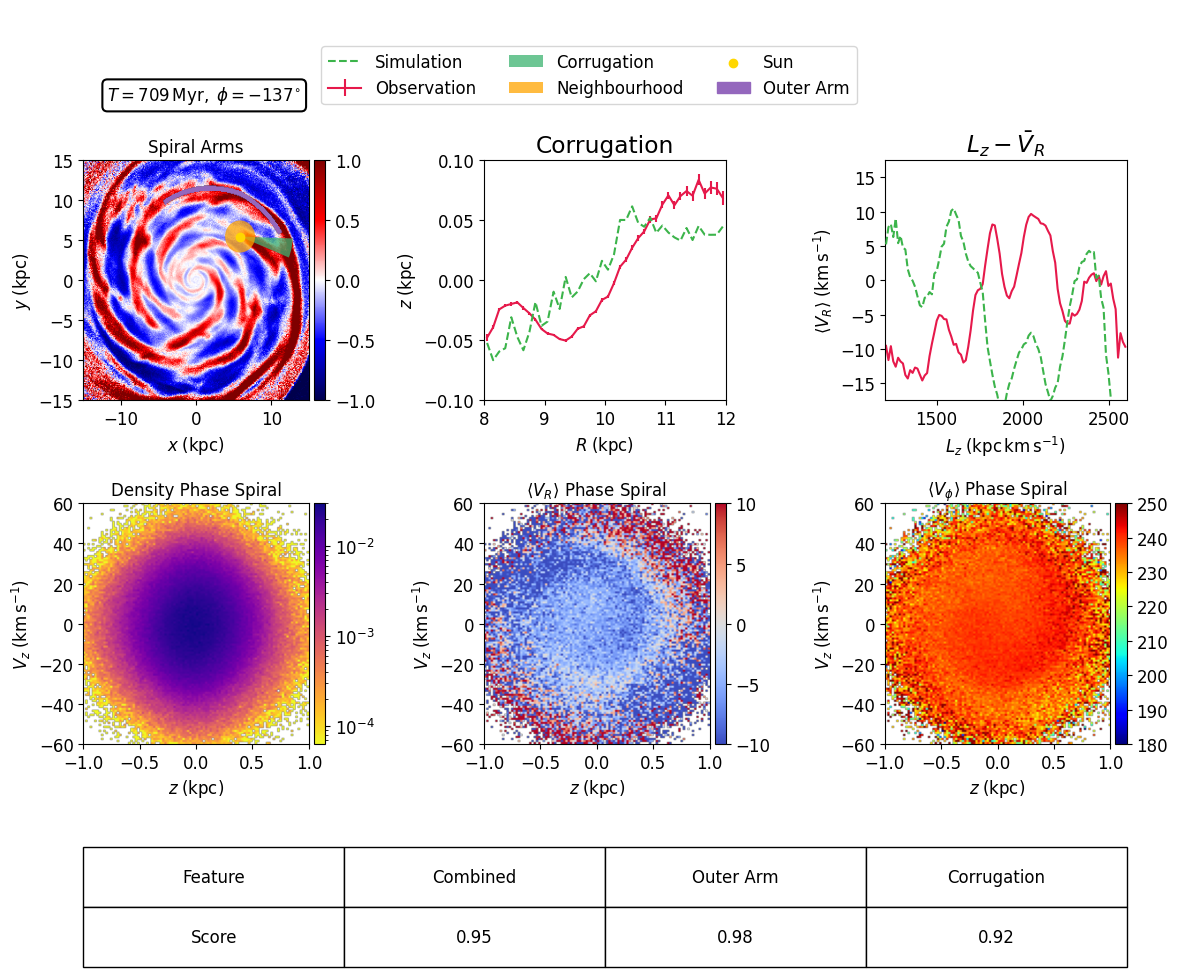}
\caption{Top left: Density contrast map of the simulated disc galaxy with an \textbf{active gas} disc. The model Outer arm is overlaid
in purple, the Sun’s position is marked in yellow, the neighbourhood used for the phase spiral is shaded in yellow, and the region used to extract the
corrugation profile is shown in green.
Top middle: Radial height profile of the disc corrugation, comparing the observed MW
profile (red) with the simulated profile (green).
Top right: The mean radial velocity binned in angular momentum or the \lzvr  wave, comparing
the observed MW profile (red) with the simulated profile (green).
Bottom left: The simulated vertical phase space distribution seen in surface number density.
Bottom middle: The simulated vertical phase space distribution seen in mean radial velocity (in units of $\kms$).
Bottom right: The simulated  vertical phase space distribution seen in mean azimuthal velocity (in units of $\kms$).
All non-observational data correspond to the simulation snapshot at $T = 708~\Myr$ and
solar azimuth $\phi = -\SI{137}{\degree}$, the configuration that maximises the
combined similarity score of the Outer arm and corrugation.
A table of the similarity scores for the corrugation and the Outer arm is located below the panels.
\label{fig:hbd_10_gd11_highlight}
}
\end{center}
\end{figure*}

The plot of the best fit epoch and azimuth pair for this model considering only the combined similarity of the Outer arm and corrugation is shown in Fig.~\ref{fig:hbd_10_gd11_highlight}.
Like with the inert gas model, the density contrast plot in the top left panel shows 
a more flocculent spiral structure than even Fig.~\ref{fig:hbd_10_gd2_highlight}. The \lzvr projection is even worse of a match than before, and the phase spirals are incredibly weak.
Although the corrugation appears to match somewhat as a general trend, that is, it starts
slightly negative and then increases until it plateaus around $R=10-11~\kpc$, it has a 
less smooth profile. This is likely due to the increased turbulence from stellar feedback
from the active gas disc.

\subsubsection{Result of Including Gas}\label{sec:gas_inclusion}
\begin{figure*}
\begin{center}
\includegraphics[width=2\columnwidth]{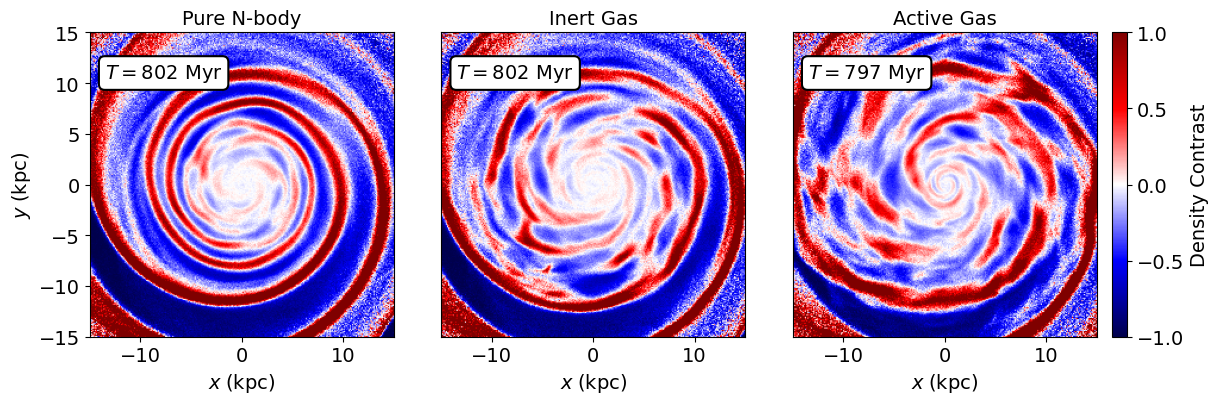}
\caption{A comparison between the density contrast of Sgr impact models at approximately the
same amount of time after the perturber's impact, $\sim 800~\Myr$. The leftmost panel
is of the pure N-body model, the middle panel is of the inert gas model and the rightmost panel
is of the star forming gas model. It is clear that including gas and then also including stellar
formation leads to more flocculent spirals.
\label{fig:all_models_compare}
}
\end{center}
\end{figure*}

As explained by \citet{ThorPhaseSpiral2025}, the inclusion
of additional gas serves to introduce a few effects:
\begin{enumerate}
\item{Increase in mass - As the stellar components are
kept identical between the models, the inclusion of gas
increases the total mass of the system. This decreases
the responsiveness of the galaxy to the external perturber
which is kept at the same mass.
}
\item{The speeding up of phase-mixing - The gas component will increase the
rate of phase mixing, which along with negatively impacting phase spirals, also
increases the rate of spiral density wave decoherence. This is clear in Fig.~\ref{fig:all_models_compare} where the gas models have much more flocculent spiral arms
compared to the pure N-body model at approximately the same time in the simulation.
}
\end{enumerate}

Therefore, it seems with our current models, the inclusion of gaseous components and feedback
hinder the match between simulations and observations. This suggests several possibilities: (i)
the gas phase may need magnetohydrodynamics (MHD) to help soften the previously mentioned effects, or (ii) that the self-gravity between stars is the more
dominant force compared to star-gas interactions, and that the feedback in the studied models
is too strong. In this respect, the sensitivity of the phase spiral to the noise means that it can be used to constrain feedback prescriptions.

\section{Conclusions}\label{sec:conclusion}
In this study, we proposed a null hypothesis - that a single crossing event can plausibly be the dominant
origin of specific MW substructures. As the number of possible signatures is increased, we find that the hypothesis must be rejected.

We analysed primarily a pure N-body simulation first presented by \citet{SgrImpactBHTG2021} consisting of a thin MW-like disc and a
disc-crossing point mass perturber mimicking the interaction between the MW and the Sgr
dwarf galaxy. We also analysed more sophisticated galaxy models that included either
an inert gas disc component or a star-forming, turbulent gas disc component. Here are the main findings:
\begin{enumerate}
\item{
	A single impact is able to produce MW-like disc corrugations and a MW-like
	Outer arm approximately $700-1200~\Myr$ after crossing at two opposing
	sides of the galaxy for all three models.
}
\item{
    The present simulations are not able to produce arm segments that match
    well with all observed MW spiral arm segments at the same azimuth and time.
    This is because each arm segment has its own pattern speed, as seen in their similarity
    maps, making it unlikely for multiple arms to match observations at the same point
    along the solar circle and in time.
}
\item{
    The largest compatible set of features are the MW Outer arm, Local arm and Sagittarius-Carina
    arm and the local disc corrugation. Additionally, the phase spiral is also qualitatively detectable along with these features.
}
\item{
    The \lzvr wave signature in our simulations does not appear to match well with observation, indicating that a single Sgr-like impact may not be responsible for this phenomena. This is in
    line with other studies that lean towards a more internal origin.
}
\item{
	Including gas appears to worsen the comparison to observation likely due to the
	increased mass dampening the effect of the satellite's perturbation and also due to
	star-gas interactions serving to decohere structures over time.
}
\item{
	Inner disc structures such as the Norma and Scutum-Centauri arm segments are not
	well reproduced in the simulations, likely due to the lack of a bar.
}
\item{
    Overall, a single impact by a satellite like Sagittarius cannot account for all disc
    substructures in the Galaxy. Future models will need to include a more sophisticated treatment of the gas, with a close examination of the non-linearity of bar-spiral arm interaction \citep{minchev12}, inter alia.
    
}
\end{enumerate}

\section*{Acknowledgements}\label{sec:acknowledgements}
We would like to thank Austin Hinkel for their assistance and clarifications on the
query and cuts used to obtain the local corrugation profile.
We are also grateful to Mark Reid for their help with regard to observational spiral
arm data.

\section*{Data Availability}
The data associated with this article is available to be shared on request.

\clearpage
\bibliographystyle{mnras}
\bibliography{references}

\appendix

\section{Spiral Arm Toy Model and Similarity Maps}\label{sec:spiral_arm_toy_model}
In order to better understand the similarity map for the spiral arms,
let us first consider a simple toy model. In this toy model we will assume the galaxy has two spiral arms
in the shape of two log spirals, one arm and its counter arm 180 degrees away.

\begin{figure}
\begin{center}
\includegraphics[width=\columnwidth]{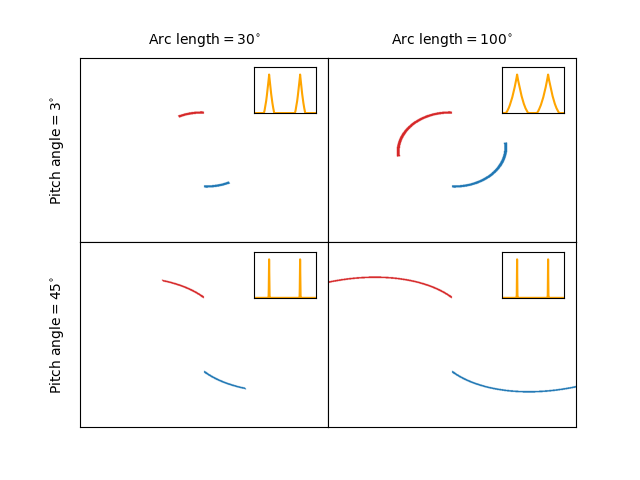}
\caption{
Each panel shows an arm and its counter-arm in red and blue along with a plot of its
coverage with respect to an identical mask arm for angles between $-{180}^{\circ}$
and $+{180}^{\circ}$  in the top right.
The top row shows arms with pitch angle ${3}^{\circ}$ whereas the bottom row shows
arms with a larger pitch angle ${45}^{\circ}$. The left column corresponds to arms
with short arc lengths of ${30}^{\circ}$ while the right column corresponds to long
arc lengths of ${100}^{\circ}$. It can be seen that the width of the peaks is much
smaller in the case of high pitch angles (bottom panels) as compared to lower pitch
angles (top panels).
\label{fig:spiral_arm_toy_models}
}
\end{center}
\end{figure}

In the ideal case where the mask arm (the observed arm we are comparing to) is identical
to the two spiral arms (the simulated arms we are matching with observation) we expect
that the coverage will peak to $100\%$ only when the mask's azimuths is equal to
the azimuths of the two spiral arms as any other angle will leave cells missing, ignoring
any potential discretisation errors.
The width over which the coverage is non-zero is dependent on the exact shape of the
spiral as the more non-circular it becomes the more difficult it is for the mask to
fit to the spiral arms at non-optimal angles. See Fig.~\ref{fig:spiral_arm_toy_models}
for examples of some configurations and their coverage profiles.

In the non-ideal cases where the mask arm differs from the underlying spiral arms, we can
expect the peaks to lower and potentially shift. Additionally if the spiral arms were
modelled instead with kinked log spirals
(like in the models presented by \citet{SpiralArmMasersReid2019}),
depending on the exact configuration of the kink, the peaks may even become lopsided.

\begin{figure}
\begin{center}
\includegraphics[width=\columnwidth]
{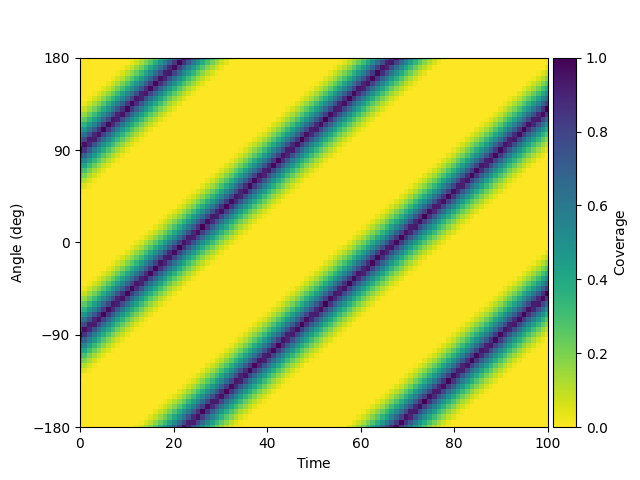}
\caption{
The coverage over time and azimuth for a system with an arm and its counter arm rotating
at a constant angular velocity. The coverage peaks where the arms are located and their
rotation leads to wrapping diagonal bands over time.
\label{fig:spiral_arm_toy_models_similarity_map}
}
\end{center}
\end{figure}

Under constant angular rotation, these peaks will travel with the same angular velocity
as the arms forming diagonal bands where the gradient is equal to the angular velocity.
As all the rotation does is impart a constant global azimuthal offset, the shapes of the
peaks will not change over time and so the diagonal bands will only have straight edges,
as can be seen in Fig.~\ref{fig:spiral_arm_toy_models_similarity_map}.

\begin{figure}
\begin{center}
\includegraphics[width=\columnwidth]
{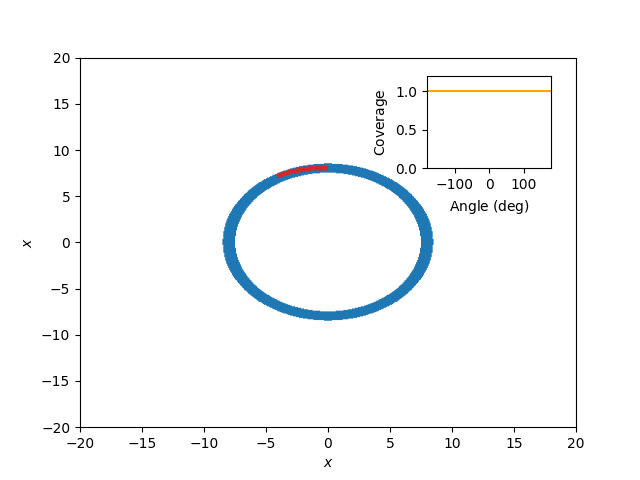}
\caption{
The degenerate case where the overdensity map contains a ring structure that encompasses
the entirety of the mask arm leading to complete coverage over all azimuths.
\label{fig:spiral_arm_toy_models_degenerate}
}
\end{center}
\end{figure}

An issue with the coverage method is that having $100\%$ coverage does not always
mean that the mask arm fits well. Any ring-like structure that covers all azimuths and
has a larger area than the mask can lead to a coverage profile that is uniformly at
$100\%$ coverage even though such a ring structure is nothing like the mask arm.
See Fig.~\ref{fig:spiral_arm_toy_models_degenerate} for an illustration.

Therefore, it is important that we see peaks and troughs in the coverage profile as that
better indicates the presence of arm segments.

\section{Sensitivity of Residual-Based Similarity}\label{appendix:residual_similarity}
The method to quantify similarity for the corrugation that used the sum of square residuals approach is more sensitive to small changes in the frequency/structure of a signal than it is to small changes in amplitude.

Let's consider an example. The observed MW corrugation radial wave can be modelled roughly as a
single sine wave of the form
\begin{equation}
    z(R) = A\sin(\omega R + \phi_{0}) + z_{0}
\end{equation}
where $z$ is the height, $A$ is the amplitude, $\omega$ is the wave frequency,
$R$ is the radius, $\phi_{0}$ is its phase and $z_{0}$ is the height at $R = 0$ or
offset.
We can imagine a slightly perturbed version of this wave to be
\begin{equation}
    z'(R) = A'\sin(\omega' R + \phi'_{0}) + z'_{0}
\end{equation}
where all parameters are slightly perturbed. We can quantify this perturbation as a relative
or normalised difference in parameters
\begin{align}
    a &= \frac{A'}{A} \\
    w &= \frac{\omega'}{\omega} \\
    \Phi &= \frac{\phi'_{0}}{\pi} \\
    Z &= \frac{z'_{0}}{A}
\end{align}
where the amplitude and wave frequency are relative to the true values, and the phase and offset
are relative to $\pi$ radians and the amplitude respectively.

\begin{figure}
\begin{center}
\includegraphics[width=\columnwidth]{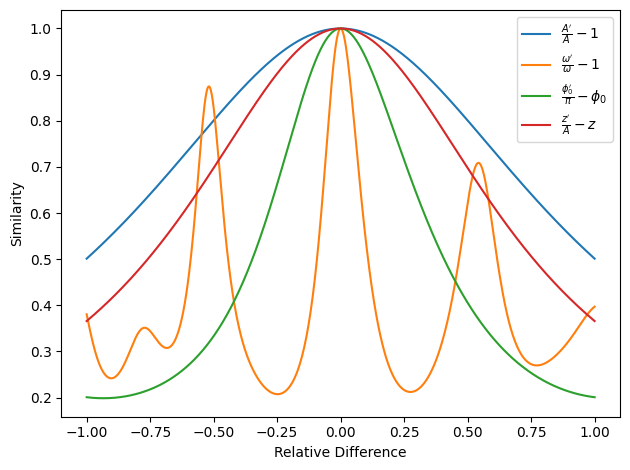}
\caption{The similarity as the amplitude, wave frequency, phase and offset are perturbed.
\label{fig:similarity_sensitivity}
}
\end{center}
\end{figure}

We can then measure how the similarity changes as a result of changing these parameters, using
the null model of a flat profile with height $0$. As can be seen from Fig.~\ref{fig:similarity_sensitivity}, amplitude does not cause much of a change in similarity
from $0$ to $2$ times the original amplitude, whereas even a small change in wave
frequency can lead to a large drop in similarity. The phase and offset are also more sensitive
than the amplitude which indicates that similarity is more sensitive to changes in the structure
of a signal that it is in the amplitude. This effect leads to more pessimistic estimations of the
similarity for features which are more complicated in structure like \lzvr  and the 2D phase spiral
signature.

\section{Derivation of Intersection Point Between Two Periodically Wrapping Lines}\label{appendix:intersection_point_derivation}
The lines cross when their azimuths match which with $\SI{360}{\deg}$ periodicity
means
\begin{equation}
	\Omega_{0} t + \phi_{0} \equiv \Omega_{1} + \phi_{1} \mod(360)
\end{equation}
where $(\Omega_{i}, \phi_{i})$ are the angular velocity and initial offset for line
$i = 0,1$.
If we write it in non-modular arithmetic notation we get
\begin{equation}
	\Omega_{0} t + \phi_{0} - \Omega_{1}t - \phi_{1} = 360 m
\end{equation}
where $m$ is some integer. Solving for time $t$ we get
\begin{equation}
	t_{\text{intersect},m} = \frac{\phi_{1} - \phi_{0} + 360m}{\Omega_{0} - \Omega_{1}}
\label{eq:intersection_time_appendix}
\end{equation}
Substituting Eqn.~\ref{eq:intersection_time_appendix} into Eqn.~\ref{eq:linear_rotation} we get
the azimuth at intersection as well
\begin{equation}
	\phi_{\text{intersect},m} = \Omega_{0}
	\left(\frac{\phi_{1} - \phi_{0} + 360m}{\Omega_{0} - \Omega_{1}}\right)
	+ \phi_{0}
\end{equation}
The variance of both coordinates can be obtained through the usual uncertainty
propagation rules.

The variance, $\sigma_{t}$ of $t$ is given by
\begin{equation}
	\sigma_{t}^{2} =
	\frac{1}{b^{2}}
	\left(
		{\sigma_{\phi_{0}}}^{2}
		+ {\sigma_{\phi_{1}}}^{2}
	\right)
	+ \frac{a^{2}}{b^{4}}
	\left(
		{\sigma_{\Omega_{0}}}^{2}
		+ {\sigma_{\Omega_{1}}}^{2}
	\right)
\end{equation}
and the variance of $\phi$ is
\begin{equation}
	\sigma_{\phi}^{2} =
	\frac{{\Omega_{1}}^{2}}{b^{2}} {\sigma_{\phi_{0}}}^{2}
	+ \frac{{\Omega_{0}}^{2}}{b^{2}} {\sigma_{\phi_{1}}}^{2}
	+ \frac{a^2{\Omega_{1}}^{2}}{b^{4}} {\sigma_{\Omega_{0}}}^{2}
	+ \frac{a^2{\Omega_{0}}^{2}}{b^{4}} {\sigma_{\Omega_{1}}}^{2}
\end{equation}
where $a = \phi_{1} - \phi_{0} + 360m$ and $b = \Omega_{0} - \Omega_{1}$.

\bsp
\label{lastpage}
\end{document}